\documentclass[12pt,preprint]{aastex}
\usepackage{amssymb}
\usepackage{txfonts}
\title{Presence of Solar Neutral Atom Corona and Coronal Heating}

\author{Z.Q. Qu$^{1*,2,3}$, R.Y. Zhou$^{4}$, H. Su$^{5}$, Y. Liang$^{6}$, L. Chang$^{1}$, X.M. Cheng$^{1}$\\
1. Yunnan Observatories, CAS, Tianwentai road, Guandu district,
Kunming, Yunnan, China\\
2. School of Space and Astronomy, Nanjing University, Hankou road,
Nanjing, China\\
3. Institute of Optics and Electronics, CAS, Shuangliu district,
Chengdu, China\\
4. Department of Physics, Chinese University of Hong Kong, Shatin,
New Territories, Hong Kong SAR, China\\
5. Yunnan Amateur Astronomers Association, No.2506 Jida Square,
Panlong district, Kunming, Yunnan, China\\
6. Zhangjiang Laboratory, Shanghai 201210, China\\
$^{*}$Corresponding author. Email: zqqu@ynao.ac.cn}
\begin{document}
\begin{abstract}
{\footnotesize By analyzing slit scanning spectral data obtained
during 2024 total solar eclipse, presence of neutral atom corona is
revealed along with a hidden inner F-corona detected within heights
below half a solar radius. The inner F-corona is found to be
essentially different from the Fraunhofer corona formed by dust
scattering beyond about 2.3 solar radius heights. The two new kinds
of the solar corona are deduced from the intensity difference
between the spectral line intensity distribution and their adjacent
continuum one acquired during the totality. Through analysis of the
recorded spectra ranging from 516.0nm to 540.7nm, the relative
depths of strong Fraunhofer lines are found to be changed among the
neighboring lines from one place to another and from the
photospheric lines acquired before the eclipse. Thus the resonant
scattering is regarded to be responsible for the changes.
Furthermore, shown in the reconstructed maps at selected spectral
lines, the detected inner F-corona is found spreading globally and
the neutral atom scattering acts as its dominant source. Both of
their distribution patterns generally depend on specific lines and
show considerably asymmetrical diffusions. The outward neutral
fluxes are observed to be mainly hindered in regions of coronal
magnetic loops with roughly estimated concentration above $10^{-5}$.
Thus it can play a key role in global coronal heating via Cowling
dissipation in these widely spread coronal loops, like heating in
Tokamak by neutral beam injection(NBI). The Cowling dissipation can
give a consistent interpretation of heating from the chromosphere to
the corona. It is suggested that the dissipation could also take
effect in magnetic reconnection regions and even trigger the
magnetohydrodynamical waves within the loops. Thus the neutral atom
corona is not only a new gradient of the solar corona but also the
missed factor critical for concatenating together these coronal
heating mechanisms.}
\end{abstract}
\maketitle

 ${\bf Key Words}$: Sun:corona, Sun: Fraunhofer corona, Sun: coronal heating

\section{Introduction}
 More than three centuries ago, Giovanni Cassini coined 'corona' for the outmost solar atmosphere after his 1706 total
solar eclipse observation. It is well known that radiation from the corona comprises spectral line emission(forming
E-corona) attributed to forbidden transitions in highly ionized ions, and continuum radiation(forming K-corona)
contributed by scattering from free electrons. Else, beyond heights of about 2.3 solar radii above solar limb, dust
grains had been found scattering the Fraunhofer(absorption) lines primitively originated in solar photosphere and
chromosphere, and an F-corona is resulted in(Grotrian,1934; Morgan and Habbal, 2007; Aschwanden, 2015; Boe et al., 2021).
However, the discovery of the F-corona had been questioned for years(Menzel and Pasachoff, 1968).

 In aspect of coronal temperature, Hardness and Young recorded the green coronal line at 530.3nm during 1869
solar eclipse, and about 65 years later Grotrian(1934) and then
Edl$\acute{e}$n(1943) realized that this spectral line is caused by
transitions in thirteen times ionized iron ions(FeXIV). And many UV
and visible emission lines with formation temperatures greater than
one million degrees have been found later during the eclipses and
via space observations. Thus a view had been accepted that the
corona should consist of only proton, highly ionized ions as well as
free electrons, with temperatures up to several million
degrees(Aschwanden, 2015). Naturally, coronal heating issue has been
raised and become a great challenge for solar physicists.

In spite of above view about solar coronal constituents, neutral
helium atoms were found in localized corona(Stellmacher and
Koutchmy, 1974; Judge and Pietarila, 2004). And an extended neutral
helium cloud in the vicinity of the Sun was detected(Kuhn, et al.,
2007; Moise et al., 2010). Else, in lower coronal layers, Bazin and
Koutchmy(2013) found two shells of neutral atoms and ions of helium.
These observations seem to strengthen a view that the solar corona
is of a multi-thermal system(Judge and Ionson, 2024) and most of the
neutral helium atoms should come from the photosphere via diffusion
across magnetic field canopies(Judge and Pietarila, 2004) rather
than from the recombination in the corona.

On the other hand, cognition of solar F-corona is more complicated
since it has been observed for a long time since its discovery in
1934(Grotrian, 1934). The most observations and derivations were
focused on its brightness and polarizations rather than its essence:
the spectral line depression below the adjacent continuum or line
depth. Ways to get these properties are indirect by measurements of
brightness and polarizations of mixed K-corona and F-corona.  The
typical works by Blackwell and Petford(1966a, 1966b), Koutchmy and
Magnant(1973) and Burtovoi et al.(2022) can represent those ways.
Meanwhile, detailed spectral measurement of the Fraunhofer lines
over a wide band in the corona during solar eclipses has had a long
history since Grotrian(1934). For instance, Allen(1946) found that
the ratio of intensity of the F-corona to that of the total
intensity(F-corona and K-corona) is independent of wavelength among
27 lines as one feature of scattering by the dust. Generally, the
Fraunhofer lines were recorded simultaneously at both neutral atom
lines and ion lines. For instance, Koutchmy and Magnant(1973)
reported the H and K Fraunhofer lines of the once ionized calcium in
form of absorption obtained by R.B. Dunn during 1965 May 30 eclipse.
High signal-to-noise observation of deep coronal spectra was
performed by Koutchmy and his cooperators(Koutchmy et al., 2009) in
the band from 510nm to 590nm. But no change in relative line
intensities and/or depths and/or the line widths among these
neighboring Fraunhofer lines were reported, as the primary feature
of the Fraunhofer lines formed by the dust scattering. Till now, the
F-corona formed by the dust has been known to have definite shape of
a hollow ellipsoid around the sun from the eclipse observation(Boe
et al., 2021) and space observations(Lamy et al., 2022; Burtovoi et
al., 2022).  Based on a great deal of data and analysis, all of
these Fraunhofer lines have been regarded to be formed via
scattering by dust for granted, lack of confirmation from the
reconstructed maps of the F-corona at specific Fraunhofer lines
respectively. It is not difficult to deduce that essential
properties of the F-corona can be revealed more directly via
Fraunhofer line depth delineation by these maps than via brightness
derivation of the F-corona under assumptions.

Using the prototype Fiber Arrayed Solar Optical
Telescopes(FASOT)(Qu, 2011; Qu et al., 2014), our polarimetry of
Fraunhofer lines in local inner solar corona during 2013 total solar
eclipse reveals existence of clouds of neutral metal atoms(Qu et
al., 2024). Stimulated from this observation, conscious explorations
revealing global distribution of the neutral atoms in the corona and
their functions in the coronal heating have been initiated.

\section{Observation and Data Reduction}

Spatial slit scanning of a Sol'Ex-V2 spectrograph improved by the
second author was performed during 2024 total solar eclipse on April
8, 2024, at Oden town, Arkansas, USA. The sky was very clear during
the totality, and thus the halo around the sun caused by the
terrestrial atmosphere can be completely ignored. This is because
that the halo brightness was estimated to be only $6.87\times
10^{-10}$ of solar disk brightness and thus two orders of magnitude
smaller than that of the F-corona near the solar limb according to
calculation of Koutchmy and Magnant(1973), and $10^{-11}$ of solar
disk brightness from calculation of Blackwell and Petford(1966a)
under the clear sky condition. The geographic coordinates of the
little town are 34.6859$^{\circ}$N latitude and 93.7760$^{\circ}$
longitude. A Schmidt-Cassegrain guiding optic system with a 150mm
aperture and calibrated focus ratio F/6.3 is used for the
spectrograph, whose slit height covers about two solar diameters and
slit width was set to be about 10$\mu m$. The dispersion recorded on
the detector chip reads as 0.012nm per pixel. The exposure time was
set to be 0.2 second. These setting parameters means that the
contribution from the dust scattering around the sun forming the
dust F-corona is comparably small after comparison with those
settings to obtain it, e.g., by Koutchmy et al.(2019). The entire
one-way scanning containing 731 spectrum frames took 2.85 minutes or
a cadence of 4.2 frames per second was performed. The spatial
extension between the two neighboring successive exposures was about
3.2 arcseconds along the scanning direction. The CMOS detector owns
a chip containing an array of 4164$\times$2116 pixels, while the
solar disk diameter covers about 2300 pixels among 4164 pixels along
slit direction . The useful observational band spans from 516.0nm to
540.7nm after the spectral line curvature correction. Three groups
of iron spectral lines are selected below to be proxies tracing the
Fraunhofer line scattering, since the iron lines are the most
abundant in the visible solar spectrum.

Three sample slit spectral images are plotted in top row of Figure
1. On the top-left is the typical photospheric and chromospheric
Fraunhofer spectral image as a reference. The top-middle and
top-right spectral images obtained during the totality are depicted
of respectively the 44th slit off the limb and the 629th slit across
the blocked solar disk. In order to see the weak Fraunhofer lines,
an artificial contrast adjustment via depressing the green coronal
line brightness is applied to the eclipse images. The rapid decrease
with height in the Fraunhofer line intensity and the continuum
background along the slit away from the middle bright region
specifies that the observed Fraunhofer lines must be primarily
originated from the scattering in the solar corona, rather than
telluric atmosphere which would smear the spatial variation along
the slit.

Correspondingly, three panels below the top row in Figure 1 give the
spectral profiles normalized by their adjacent continua, drawn
respectively from the three samples. They are obtained after
integration over the coronal loop regimes(refer to the bottom right
panel of Fig.3 below) indicated by dashed lines and outside them
indicated by solid lines producing the Fraunhofer lines. Only lower
part of the green coronal line profile is shown here for
highlighting the Fraunhofer lines in the lowest two eclipse
profiles.

\begin{figure}
\flushleft\vspace{-0.8cm}\hspace{0.26cm}
\includegraphics[width=5.0cm,height=7.6cm]{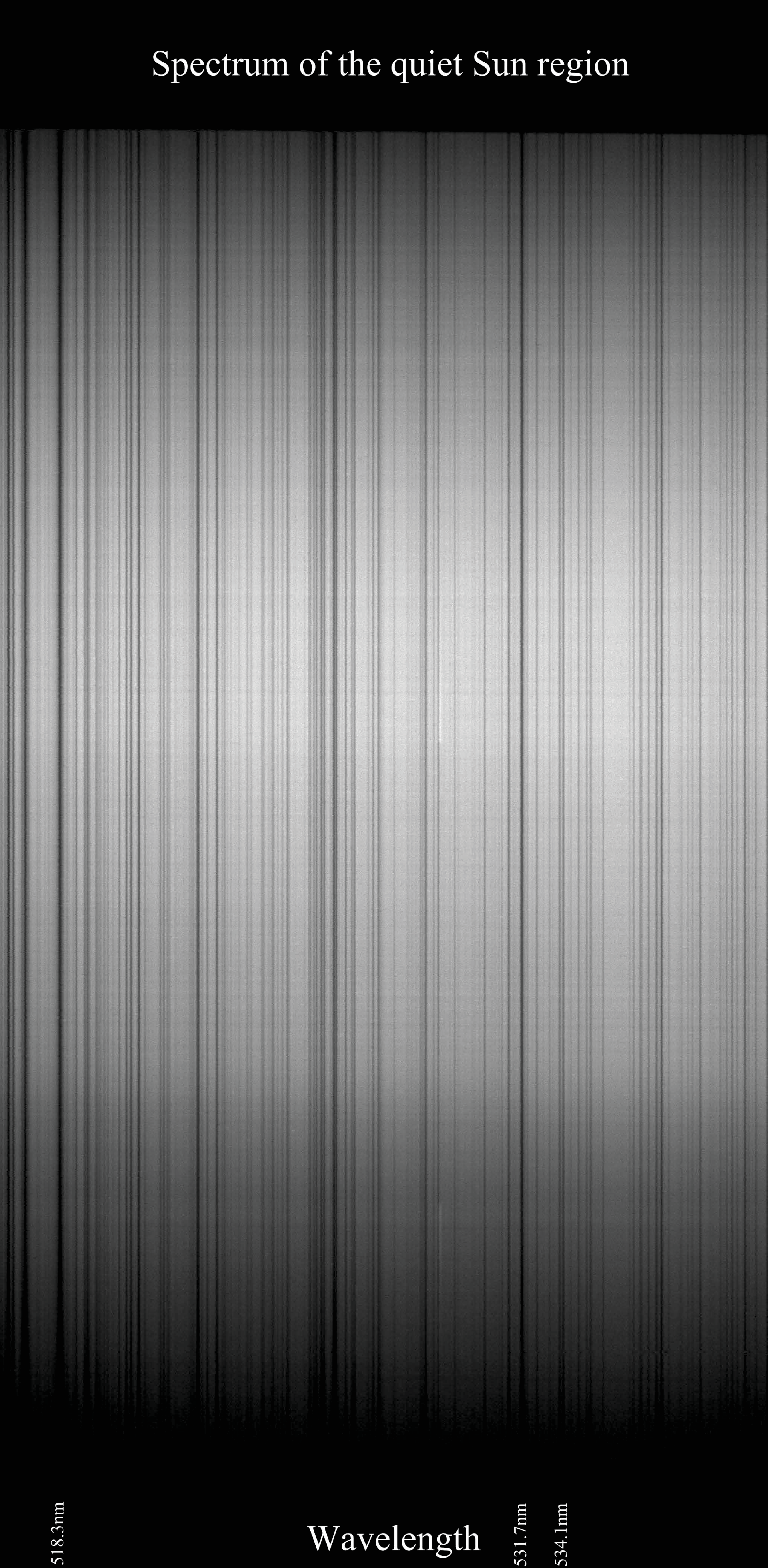}
\includegraphics[width=5.0cm,height=7.6cm]{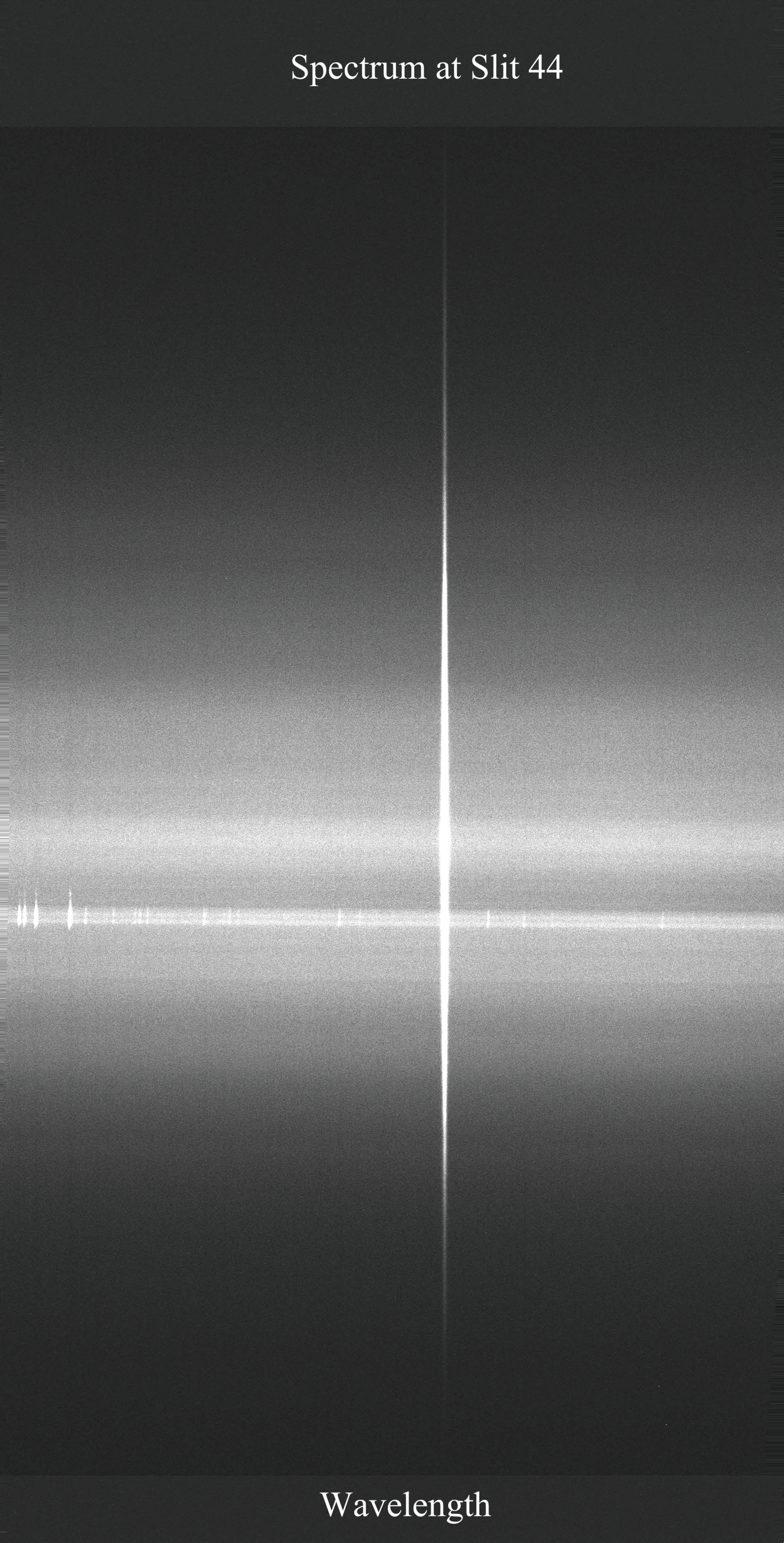}
\includegraphics[width=5.0cm,height=7.6cm]{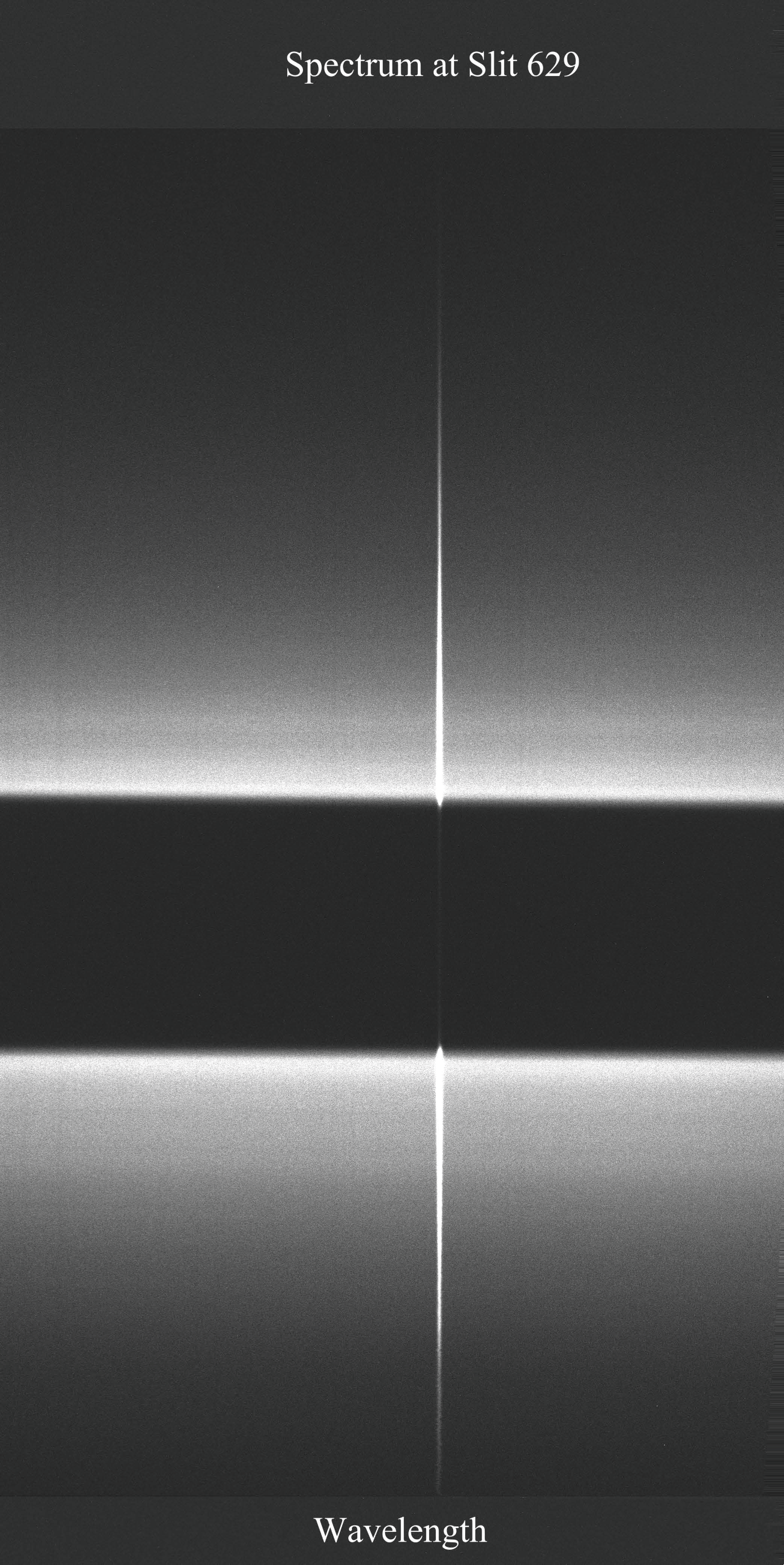}\\
\centering\hspace{-2.6cm}
\includegraphics[width=19.0cm,height=4.6cm]{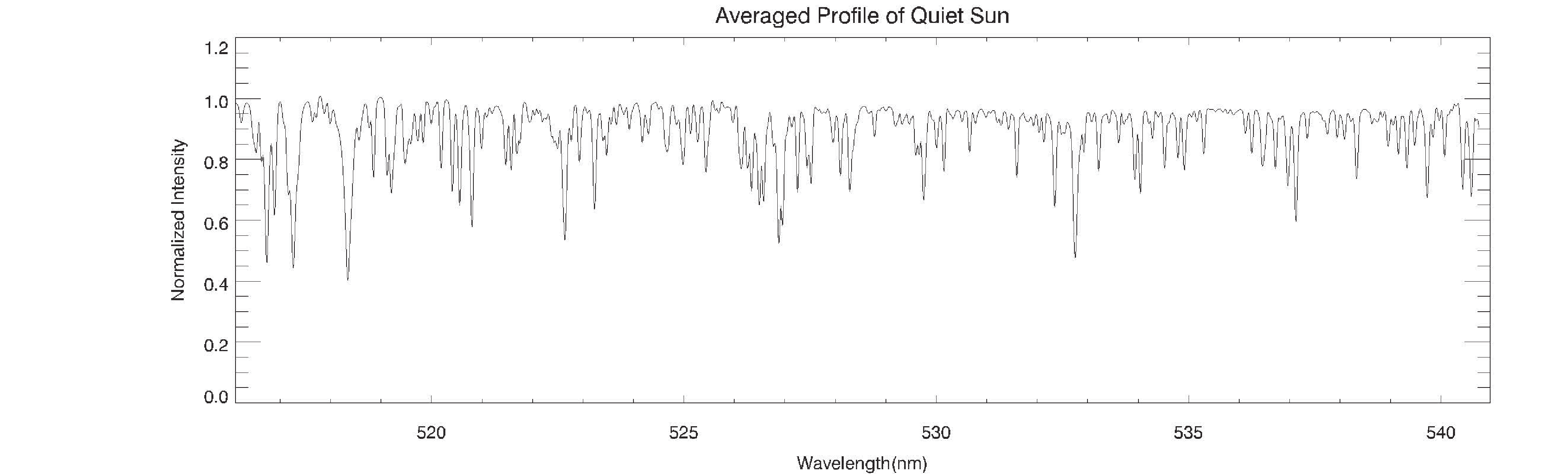}\\
\centering\hspace{-2.6cm}
\includegraphics[width=19.0cm,height=4.6cm]{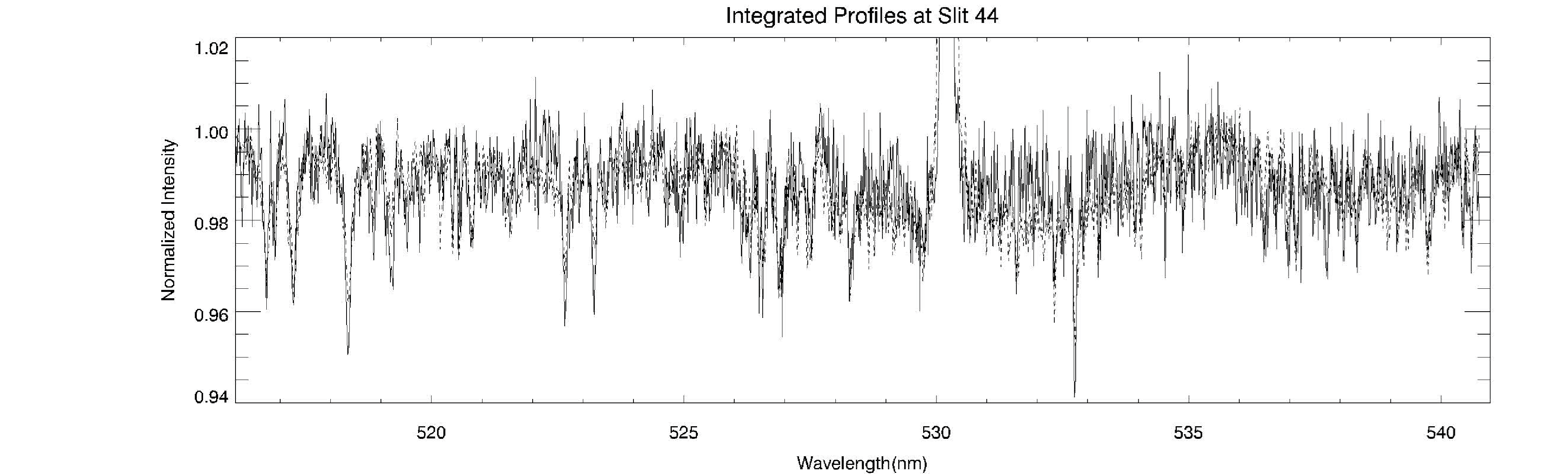}\\
\centering\hspace{-2.6cm}
\includegraphics[width=19.0cm,height=4.6cm]{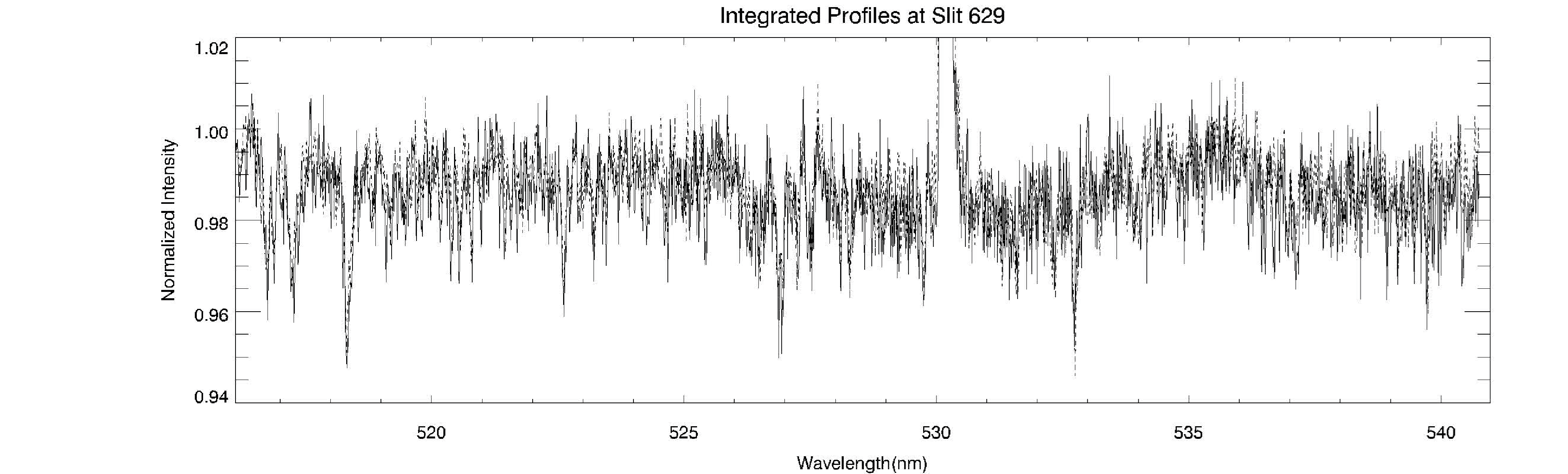}
\caption{\footnotesize Sample spectral images and induced profiles.
The top-left image was acquired with the slit across the bright
solar disk before the total eclipse as a reference. Images in the
top-middle and top-right panels were obtained respectively at the
44th and 629th scanning slit positions during the totality. The
corresponding spectral profiles derived after spatial average over
the slit pixels acquired about the coronal loop regions(dashed
lines) and outside the loops(solid lines) are plotted in the lowest
two panels, where only far wings of the FeXIV 530.3nm profile are
shown to highlight the Fraunhofer lines.}
\end{figure}

In order to improve the signal-to-noise ratio, binning of four
pixels into one along the slit is applied. The line depth and
emission amplitudes, yielding respectively maps of the F-corona and
E-corona, are emphasized via the intensity difference obtained from
subtraction of the integrated spectral line intensity from their
adjacent continuum one. The subtraction result can be either
negative reflecting the scattering-induced Fraunhofer line
depression or positive representing the line emission by their usual
meanings.

Finally, the measurement noise described by the standard deviation
$\sigma$ is estimated from the completely occulted disk, since
intensities on the disk would be extremely close to zero during the
totality.

\section{Detection Results}

It is easy to find in the lower three profile panels of Fig.1 that
profile variations in relative line depth and width are beyond noise
level. One of the most striking change from the reference profile of
the quiet sun occurs in the relative depths of three Fraunhofer
lines between 520.3nm and 521.0nm, that they become comparable to
each other while increase with wavelength can be seen in the
reference. Another considerable variations can be witnessed in the
relative line depths among the neutral magnesium triplet (MgI
$b_{1}: 518.4nm, b_{2}: 517.3nm, b_{4}: 516.7nm$), in which relative
difference in the $b_1$ line depth of about $5\%$ seems to be
enlarged when comparing to the other two of about $4\%$, while in
the reference profile, the difference are relatively smaller. It is
distinct that the two lines at about 532.8nm, yielded by the neutral
iron and chromium atoms, can become much more prominent within the
band. And so are the mixed lines at 526.9nm at the slit 629, where
the two lines become more separate to each other. The variations in
different regions can be also found between the two eclipse spectra,
which means also the variation from the reference profile. For
instance, the line depth and width of lines at 519.2nm and 523.2nm
are greater in the region without detectable loops(solid line) than
that region around the loops(dashed line) at the slit 44, and also
than those at the slit 629. Else, differences of the relative depths
of lines at 526.5nm and 526.9nm are enlarged from the slit 44 to the
slit 629, but vice versa for the mixed lines at 528.3nm. In fact, in
different height regions at the same slit, both the relative line
depths and widths can be also variant. For example, at the slit 44,
the relative depths of the strong lines integrated in the lower
regions(represented by dashed line) are generally shallower than the
higher ones(solid line). But few exceptions exist, e.g., for lines
at 520.2nm, 523.3nm, 527.3nm, 528.7nm, 532.4nm and 539.7nm. The line
width difference can be easily found at the MgI$b_1$, 519.2nm,
522.7nm, 523.3nm and even at the green coronal line. The width
variation can be ascribed to variation in combination of temperature
and line-of-sight macro turbulent velocity. Similarly at the slit
629, the variation in the relative line depth is evident at $b_1$,
$b_2$, $b_3$, 519.2nm, 520.2nm, 520.9nm, 524.2nm, 525.3nm, 526.9nm,
527.0nm, 532.8nm, 536.7nm and 540.4nm lines. Otherwise, the
variation in the line width can be clearly observed of the $b_1$,
$b_2$ and $b_3$ lines, and lines at 520.5nm, 528.3nm, 529.8nm,
532.3nm and 532.8nm.

The above variations specify enough that scattering by most of the
strong Fraunhofer line in the corona detected here depends on not
only the spatial positions but also specific spectral lines resulted
from individual atoms and ions with different ionization and
excitation states rather than bound in the dust grains(cf.
Russell,1929). In other words, it is reasonable to deduce that
resonant scattering should take place at those changed strong lines.
Correspondingly, the relative variations among those lines like the
magnesium Fraunhofer triplet are resulted from the atomic structure
and local physical environment integrated along the line-of-sight.
In more detail, the scattered intensities depend on not only their
oscillator strengths but also the localized temperatures, as shown
in Eq.(2) below. Definitely, such relative line depth variations
exclude their origin from telluric atmospheric Rayleigh scattering
of solar photospheric radiation. This is because the Rayleigh
scattering is proportional to $\lambda^{-4}$, which cannot change
the relative line depths of the photospheric lines, as the striking
advantage of the total solar eclipses(Koutchmy et al., 2019).

Another more stronger evidence supporting the above view lies at the
spatial distribution of the Fraunhofer line scattering intensity,
shown in reconstructed maps of iron lines in Figure 2. The left
column is managed for the FeI526.9nm line, the middle for the
FeI534.1nm, and the right for the FeII531.7nm. The top row maps
present distributions of the integrated line intensities in unit of
detector readouts proportional to photon flux received. Evidently,
the radiations are more concentrated in the low latitudes, and more
radiations in the south than the north. This cannot be replicated as
the solar halo resulted from scattering by the telluric atmosphere.
The directions are indicated in the black disk of the top-middle
panel. Below the top row are maps of the differences between the
line intensity and its adjacent continuum, in form of respectively
negative values(i.e., line intensity below the continuum)
representing the Fraunhofer line depression(middle row) and positive
values(i.e., line intensity above the continuum) indicating the line
emission(bottom row).

These maps in the middle row are called 'monochromatic' F-corona at
the sample lines. The left and middle maps present the specific
neutral atom F-coronae while the middle-right map is related to the
once ionized iron ion at 531.7nm. The F-corona maps in the middle
row panels show us both the common features and differences in these
derived Fraunhofer line scattering distributions. The striking
common features can be seen that the scattering is globally
detectable, and the Fraunhofer line depressions show an overall
divergence with height that signifies scatterer diffusion.
Furthermore, no observable inner fine structures like fibrils or
loops can be found unless they are shaped by the coronal loops.
Critically, different asymmetric distribution patterns are
simultaneously seen in all these three scattering distributions. For
instance, all of these distributions show outstanding east-west and
north-south asymmetries but the patterns are different. The
distributions at both the FeI526.9nm and FeII531.7nm maps display
the asymmetries that the east-west and north-south asymmetries are
much weaker than the FeI534.1nm map. However, detailed view tells us
that the scattering by the neutral atoms is more concentrated in the
low latitudes in the FeI526.9nm map than in the 531.7nm one.  As
indicated by the corresponding grey scale vertical bars in the left
sides of these maps, the line depressions are on the whole a little
stronger at the once ionized iron ion line respectively than the
neutral atom lines plotted here, the line depressions from the
neutral magnesium triplet and neutral iron line at 527.0nm(not
depicted here) are respectively overall thicker than the 531.7nm.
Since much less ion lines can be detectable within the band and the
FeII531.7nm line ranks the strongest, the dominant contribution to
the F-corona is from the neutral atoms(refer to the bottom left
panel of Fig.3). In summary, the most important thing found is that
the line depression spatial distribution patterns and strengths
depend on specific Fraunhofer lines. This suggests strongly that
these atoms and ions as scatterers are separate from each other
rather than confined in the dust grains.

It is noteworthy that signals of the depressions just around the
disk and within the loops can be well above the noise level,
evaluated by the corresponding $\sigma$: 1.48$\times 10^{3}$ readout
units for the FeI526.9nm line, 5.13$\times 10^{3}$ readout units for
the FeI534.1nm, and 2.24$\times 10^{3}$  units for the FeII531.7nm
lines. Evidently, the heights extended by the neutral atom corona
and F-corona detected here are below half solar radius accordingly
in this measurement.

All of above observational facts specify presence of the neutral
atom corona, and the F-corona described here is essentially
disparate from that F-corona formed by the dust grains scattering
the photospheric and chromospheric radiation above heights of about
2.3 solar radii. Actually, the dust grains will be sublimated due to
solar heat and radiation below these heights(Russell,1929).
According to space observations(Lamy et al., 2022; Burtovoi et al.,
2022), this 'outer' dust F-corona has an ellipse-shaped distribution
around the sun, symmetric about the elliptic axes and extended to
the earth orbit, seen as the Zodiacal light. Therefore, the F-corona
detected here is dubbed 'inner F-corona'.

In fact, the neutral atoms in the corona contribute also to the line
emission, as witnessed in the bottom two profiles in Fig.1, e.g.,
neutral iron lines at 517.9nm, 522.2nm, 524.4nm, 532.4nm and 539.9nm
at the slit 44, and at 517.6nm, 522.3nm, 527.3nm and 530.7nm. The
excitation potentials of upper energy levels causing these lines
rank from 3.879$eV$ for FeI517.6nm to 5.033$eV$ for FeI527.3nm
according to
NIST($https://physics.nist.gov/PhysRefData/ASD/lines$\_$form.html$),
much lower than those of emission lines detected by Stellmacher and
Koutchmy(1974). But our explanation is inclined to that by Deutsch
and Righini(1964) as presence of cool iron clouds. This is because
that if they were formed via Rayleigh scattering by terrestrial
atmosphere of solar chromospheric emission lines, other stronger
emission lines within this band like the magnesium triplet should be
observed simultaneously at the same region. This is not the case, as
evidenced by the bottom two profiles in Fig.1. The emission
distributions can be traced at these two representative neutral atom
lines in the bottom left and middle maps of Fig.2. In order to
clearly see the distributions, a contrast change is used via
depressing the prominence brightness in maps of FeI526.9nm and
FeII531.7nm. The emission distribution seems close to their line
depression counterparts for the FeI526.9nm and FeII531.7nm, while an
oppositely asymmetric distribution compared to its F-corona
counterpart is seen in map of the FeI534.1nm line.

\begin{figure}
\centering\vspace{0.9cm}
\includegraphics[width=5.3cm,height=3.9cm]{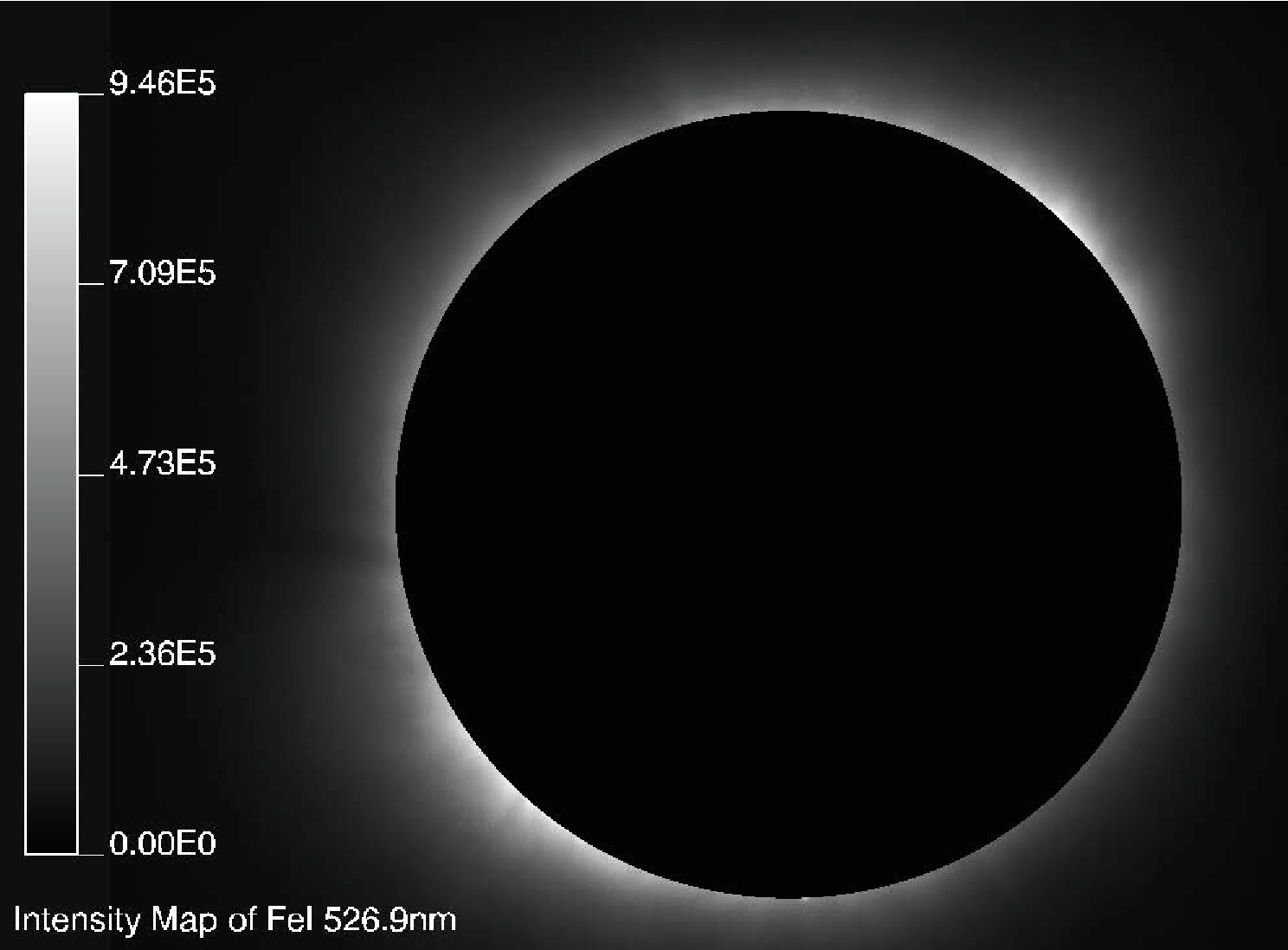}
\hspace{0.1cm}
\includegraphics[width=5.3cm,height=3.9cm]{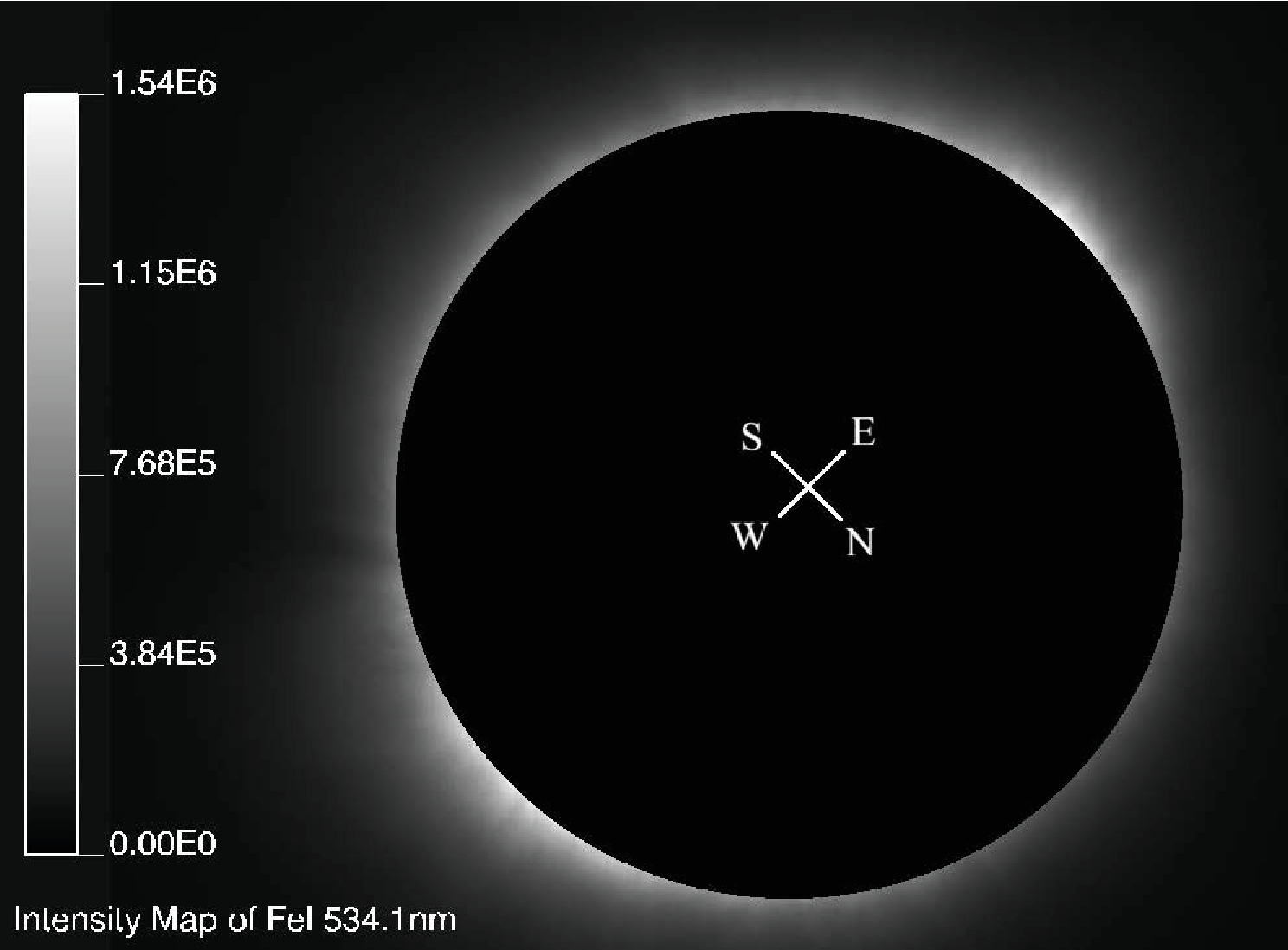}
\hspace{0.1cm}
\includegraphics[width=5.3cm,height=3.9cm]{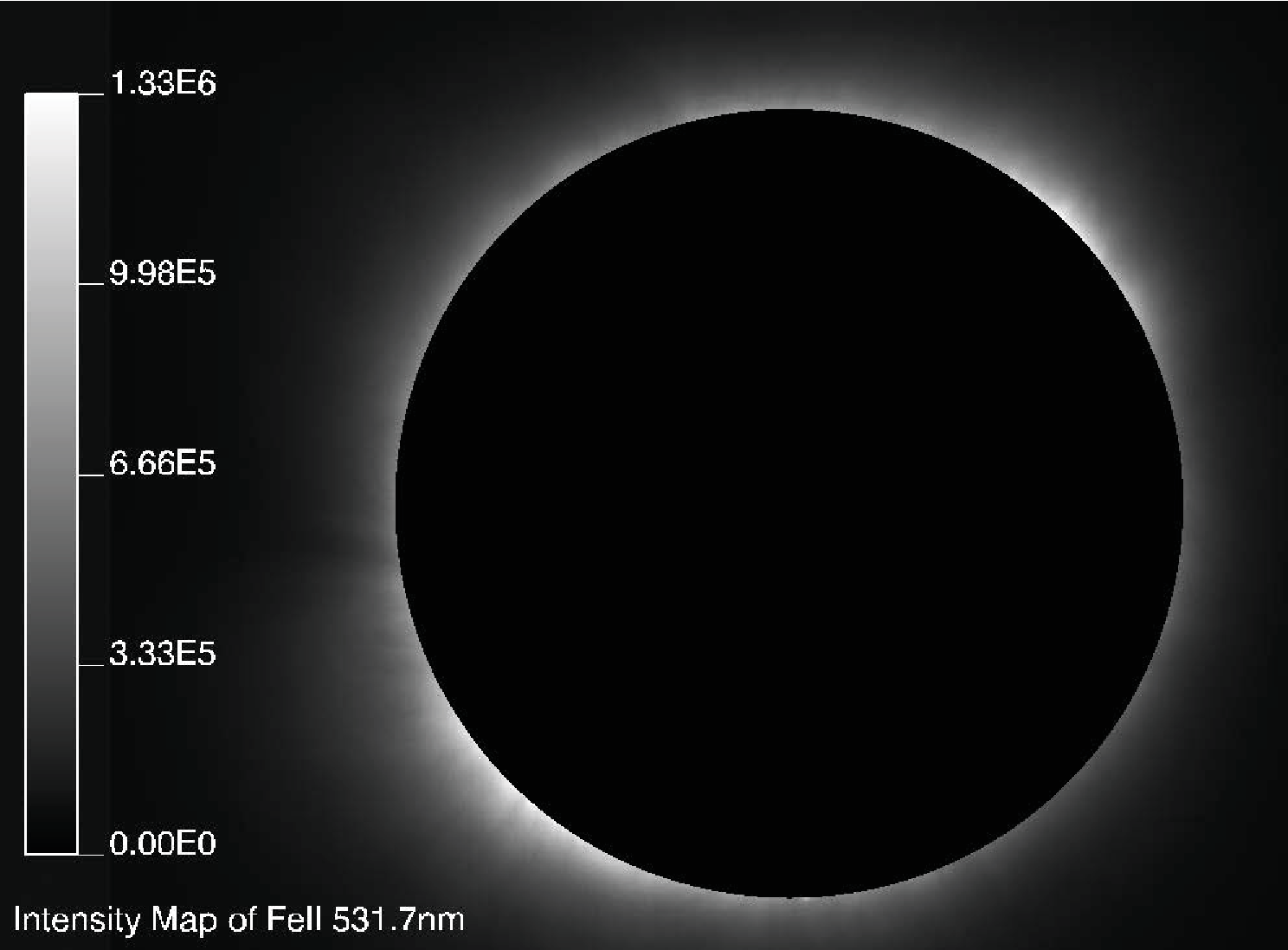}\\
\centering\vspace{0.3cm}
\includegraphics[width=5.3cm,height=3.9cm]{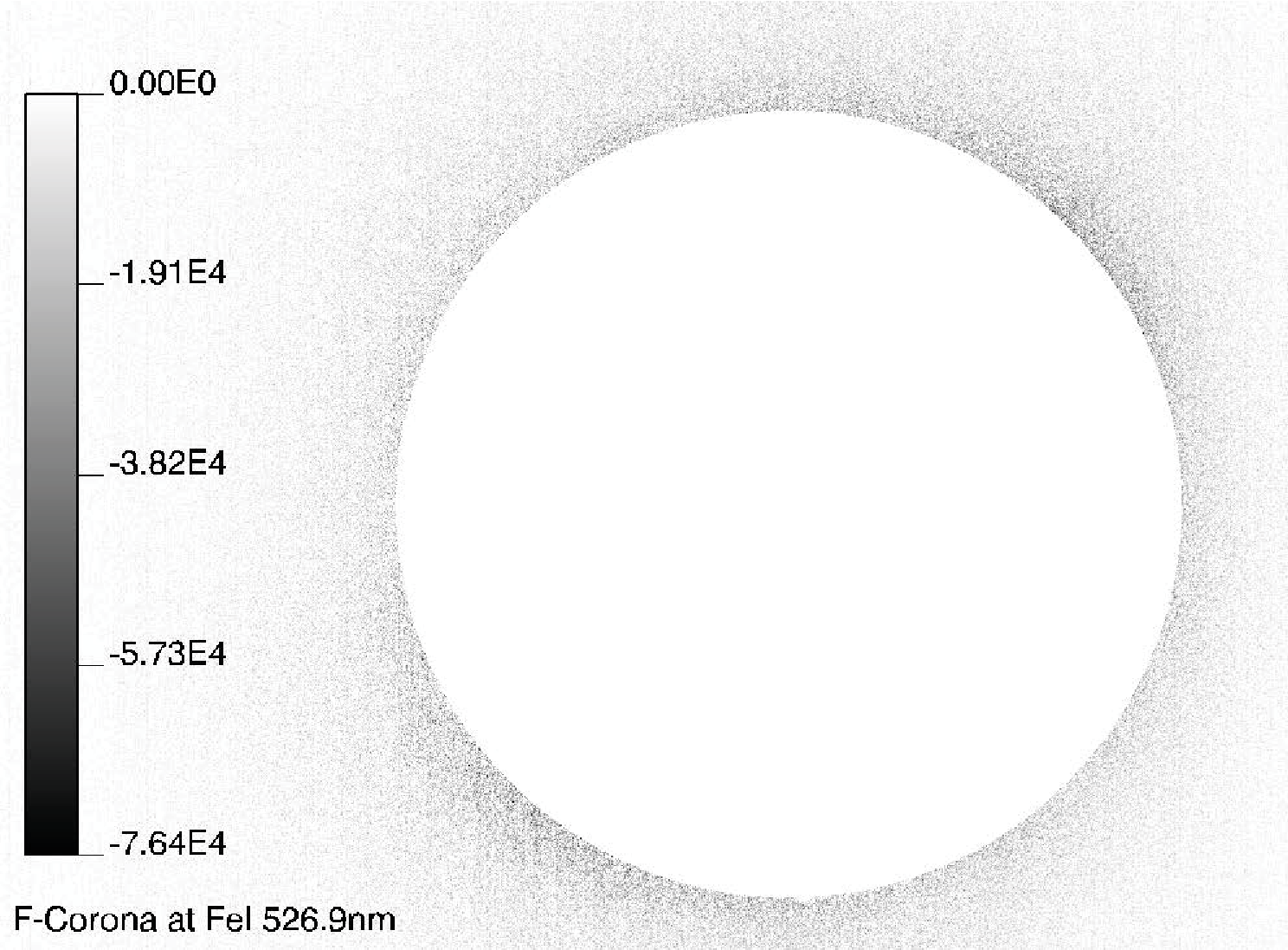}
\hspace{0.1cm}
\includegraphics[width=5.3cm,height=3.9cm]{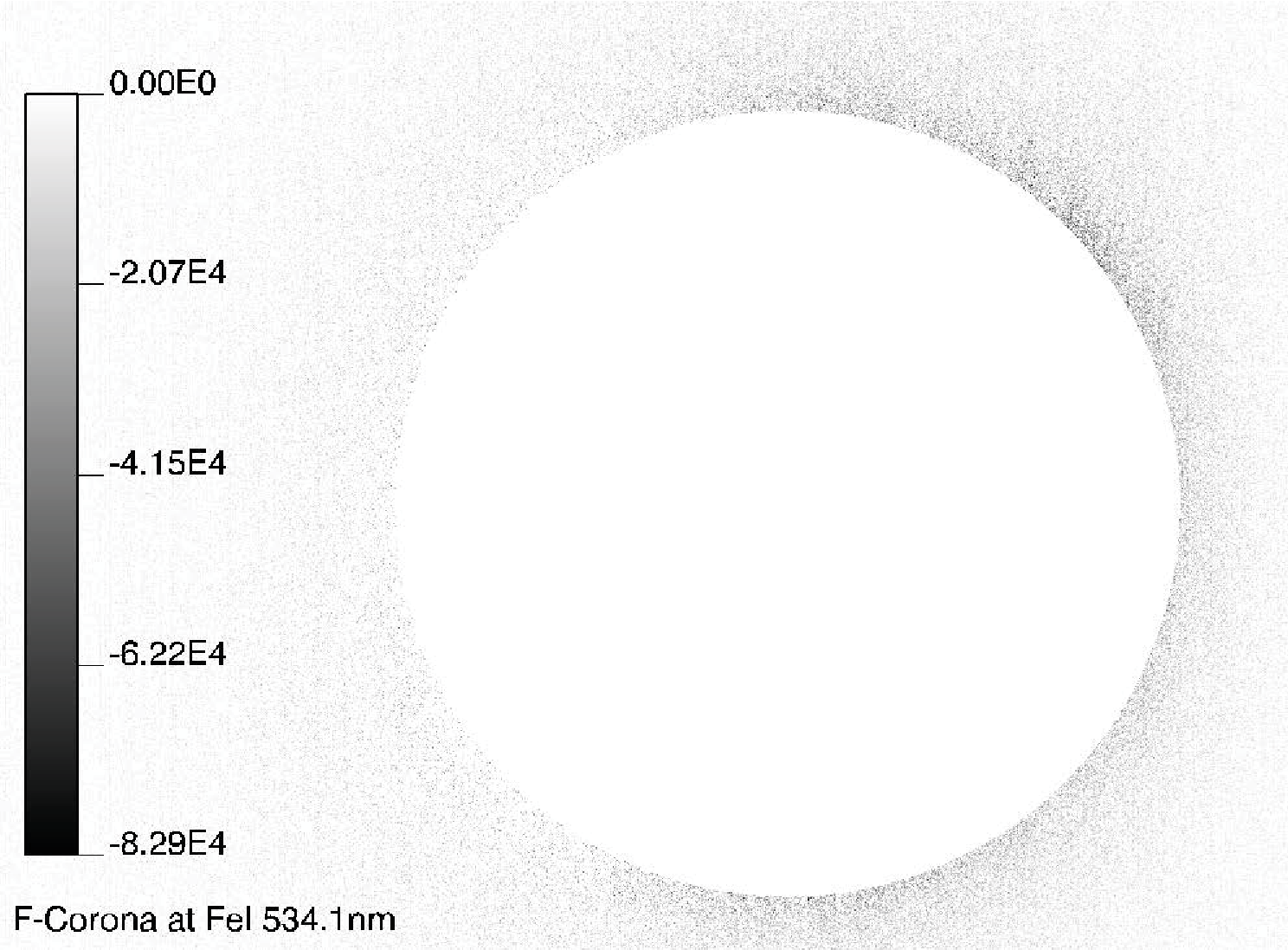}
\hspace{0.1cm}
\includegraphics[width=5.3cm,height=3.9cm]{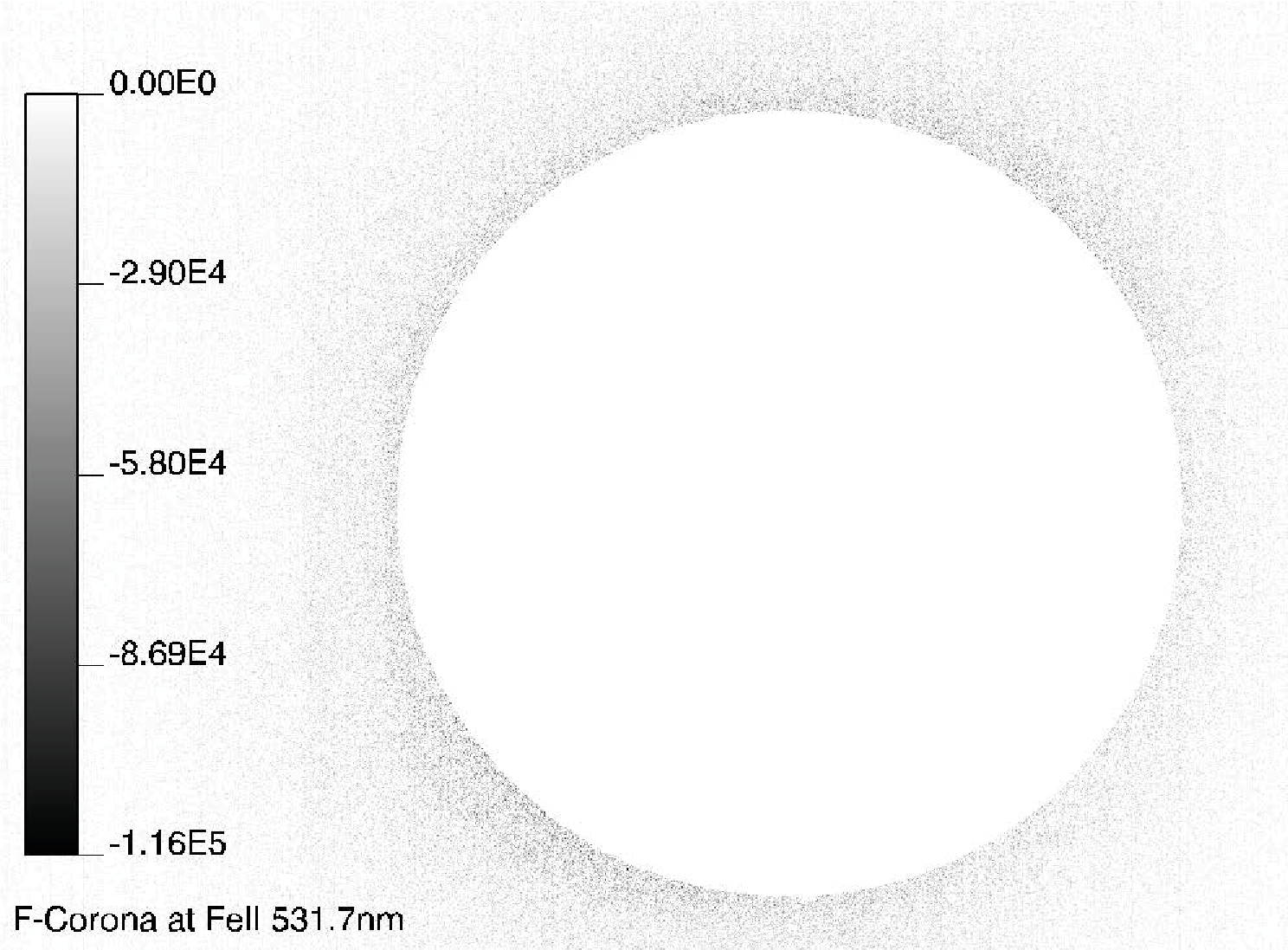}\\
\centering\vspace{0.3cm}
\includegraphics[width=5.3cm,height=3.9cm]{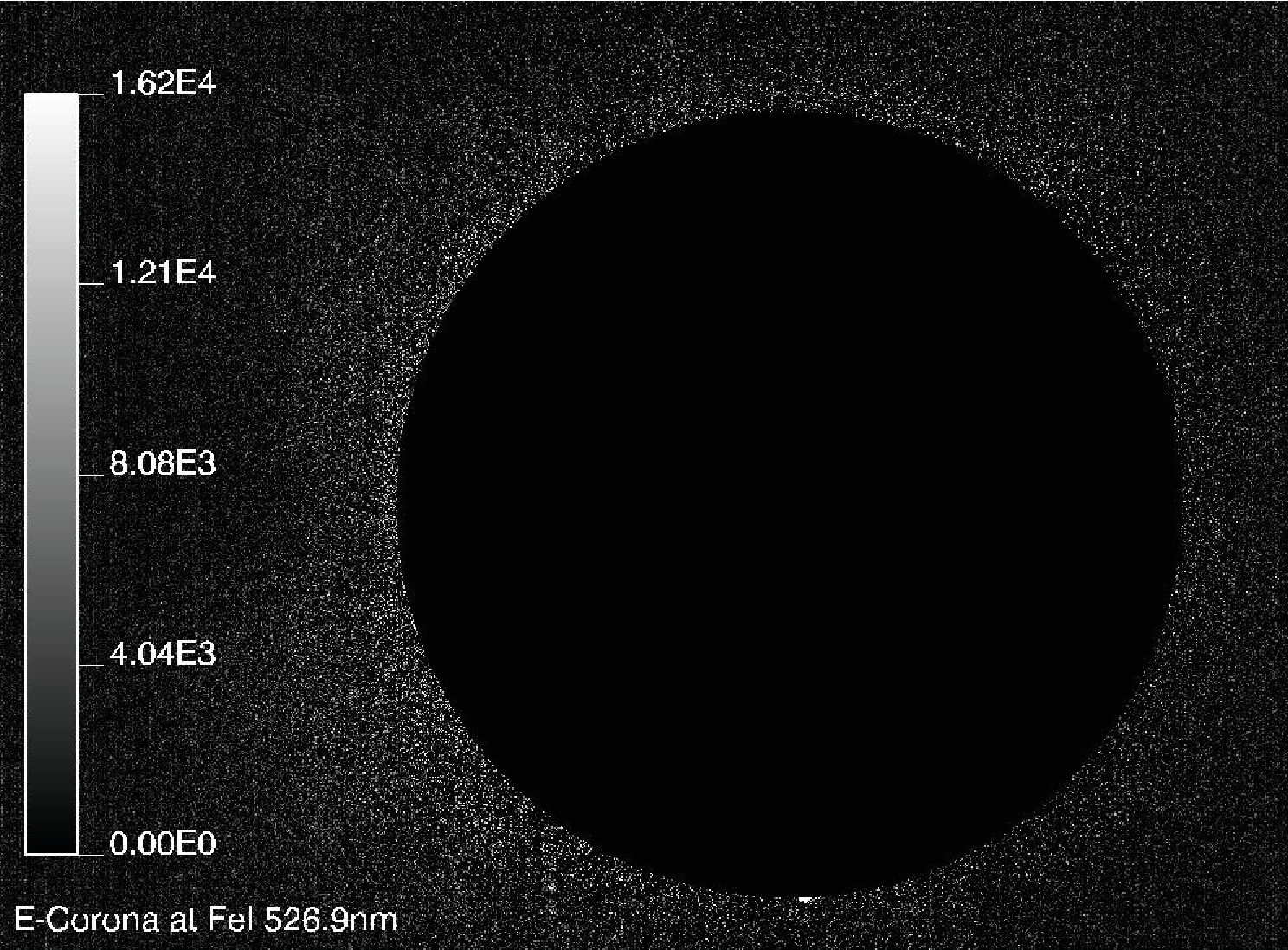}
\hspace{0.1cm}
\includegraphics[width=5.3cm,height=3.9cm]{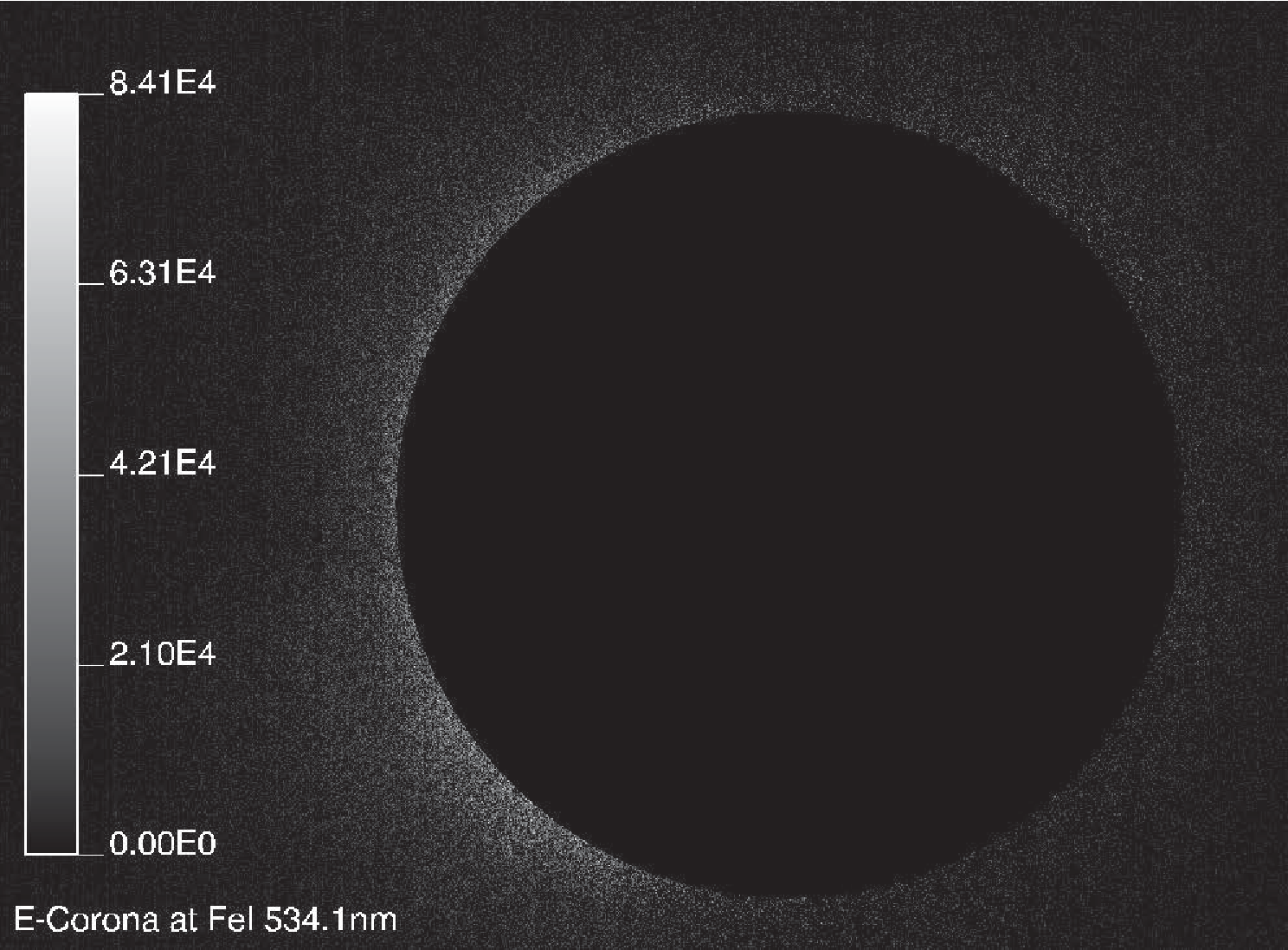}
\hspace{0.1cm}
\includegraphics[width=5.3cm,height=3.9cm]{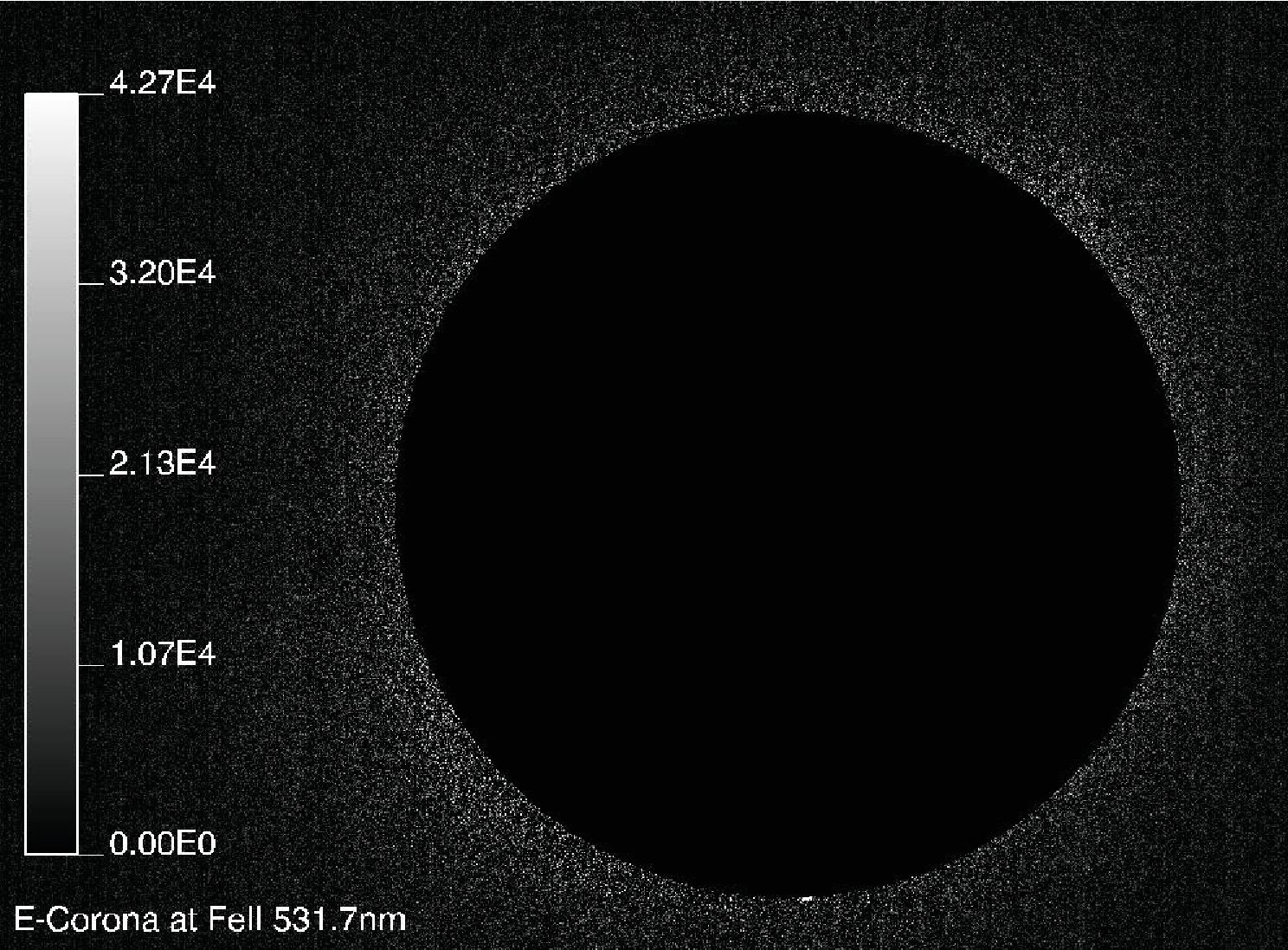}
\caption{\footnotesize Maps of spectral line intensities(top
panels), derived 'monochromatic' F-coronae(middle panels) and
derived 'monochromatic' emission(the bottom panels). They are
plotted respectively for neutral iron atom line at 526.9nm(left
column) and lines at 534.1nm(middle column), and once ionized iron
ion lines at 531.7nm(right column) for comparison. The directions
are plotted on the dark disk of the top-middle panel. Note that the
east-west and north-south asymmetries exist in the scattering
distributions of these different state iron particles on which the
distribution patterns depend. }
\end{figure}

The neutral atom corona and the corresponding inner F-corona
obtained by summation over the whole observational band are
respectively presented in Figure 3, together with corresponding
K-corona and E-corona. Directions are indicated on the occulted disk
in the bottom right panel. The neutral atom corona map shown in the
top left panel contains both the line depression(dark) and
emission(bright) contributed from different neutral atoms yielding
more than fifty lines. However, only those spatial points are
included here with magnitudes of the intensity differences larger
than three times standard deviations 3$\sigma(1.74\times 10^{5}$
readout units). It is evident that in the corona, the Fraunhofer
line depression dominates over its emission counterpart that is
spread with more scattering. It is easy to find that the north-south
and east-west asymmetries still remain.

Another way to describe neutral atom scattering distribution is
adopting absolute values of the intensity difference greater than
$3\sigma$, depicted in the top right panel. All these features of
the diffusions, non-existence of fine structures and the asymmetries
can be again witnessed in the map. The densest neutral atoms are
confined in the coronal loops clearly seen in the bottom right map.
The detectable neutral atom corona is judged here to extend mainly
to a height of half a solar radius again. The corresponding K-corona
depicted in the middle left panel as the denominator is used to
provide a fractional intensity difference magnitude map of the
neutral atom scattering at each spatial point, shown in the middle
right panel. In fact, such a map of the neutral atom scattering
gives distribution of the neutral atom concentration relative to the
free electron concentration, as is estimated in the next section.

The F-corona above 3$\sigma(1.29\times 10^{5}$ readout units),
contributed from more than fifty neutral atom Fraunhofer lines and
thirty ion Fraunhofer lines, is plotted in the bottom left panel. It
is seen that the deepest line depths lie in the loops and all the
common distribution features derived from the monochromatic ones
survive, similar to the neutral atom corona. Once again, they must
be of solar corona and exclude completely their origin of the
telluric atmosphere.

The strong correlation of the neutral atom corona depicted in the
top right panel and the inner F-corona in the bottom left panel with
distribution of the coronal loops shown in the bottom row of Fig.3
specifies that it differs critically in aspect of morphology from
the 'outer' dust F-corona with shape of hollow ellipsoid. On the
other hand, the correlation illustrates that the neutral atom
outwards diffusion is impeded by the loops which contains ions and
free electrons forming the electric current. Thus there should be
something happening.

\begin{figure}
\center
\includegraphics[width=7.2cm,height=5.3cm]{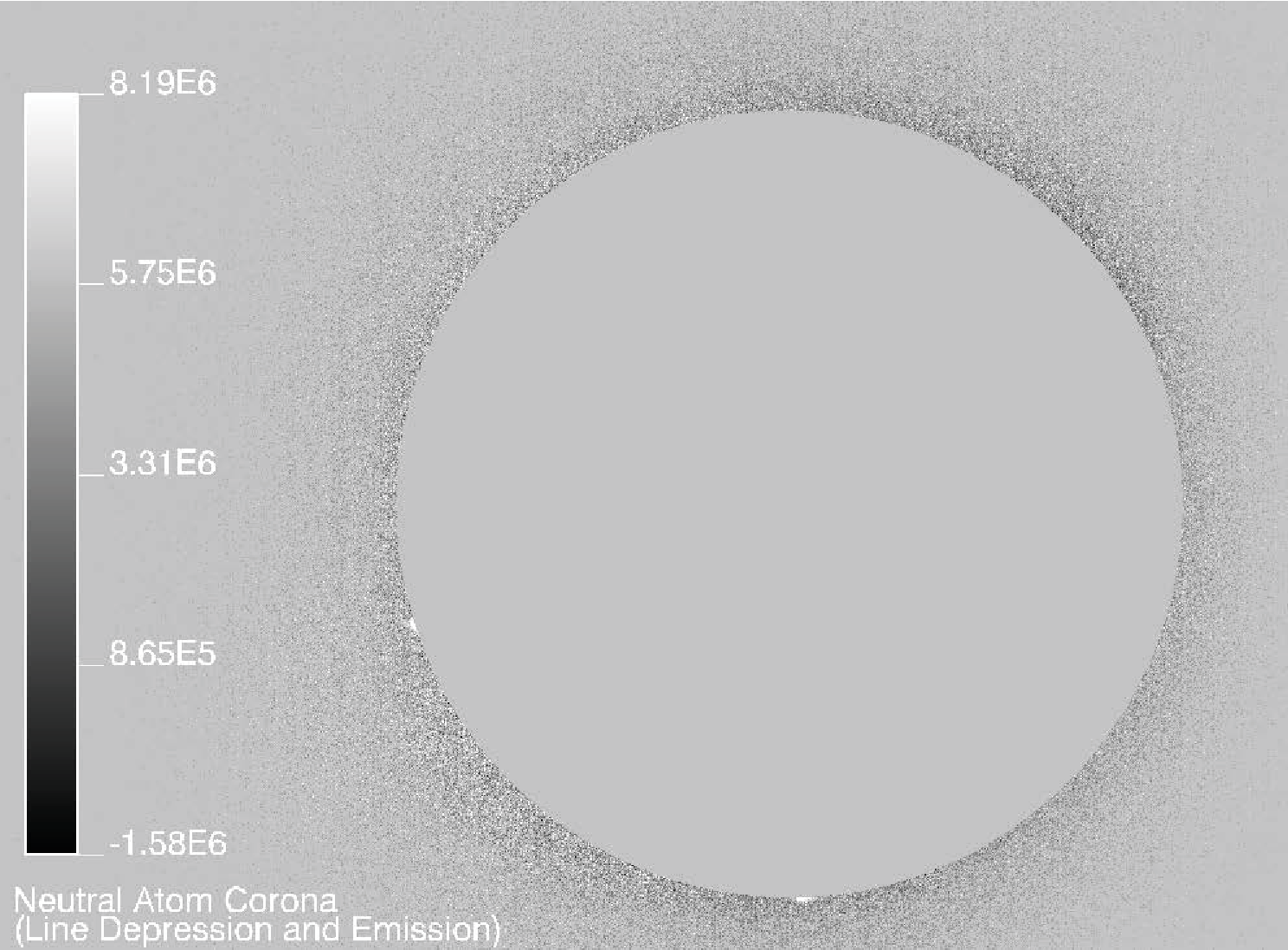}
\hspace{0.2cm}
\includegraphics[width=7.4cm,height=5.3cm]{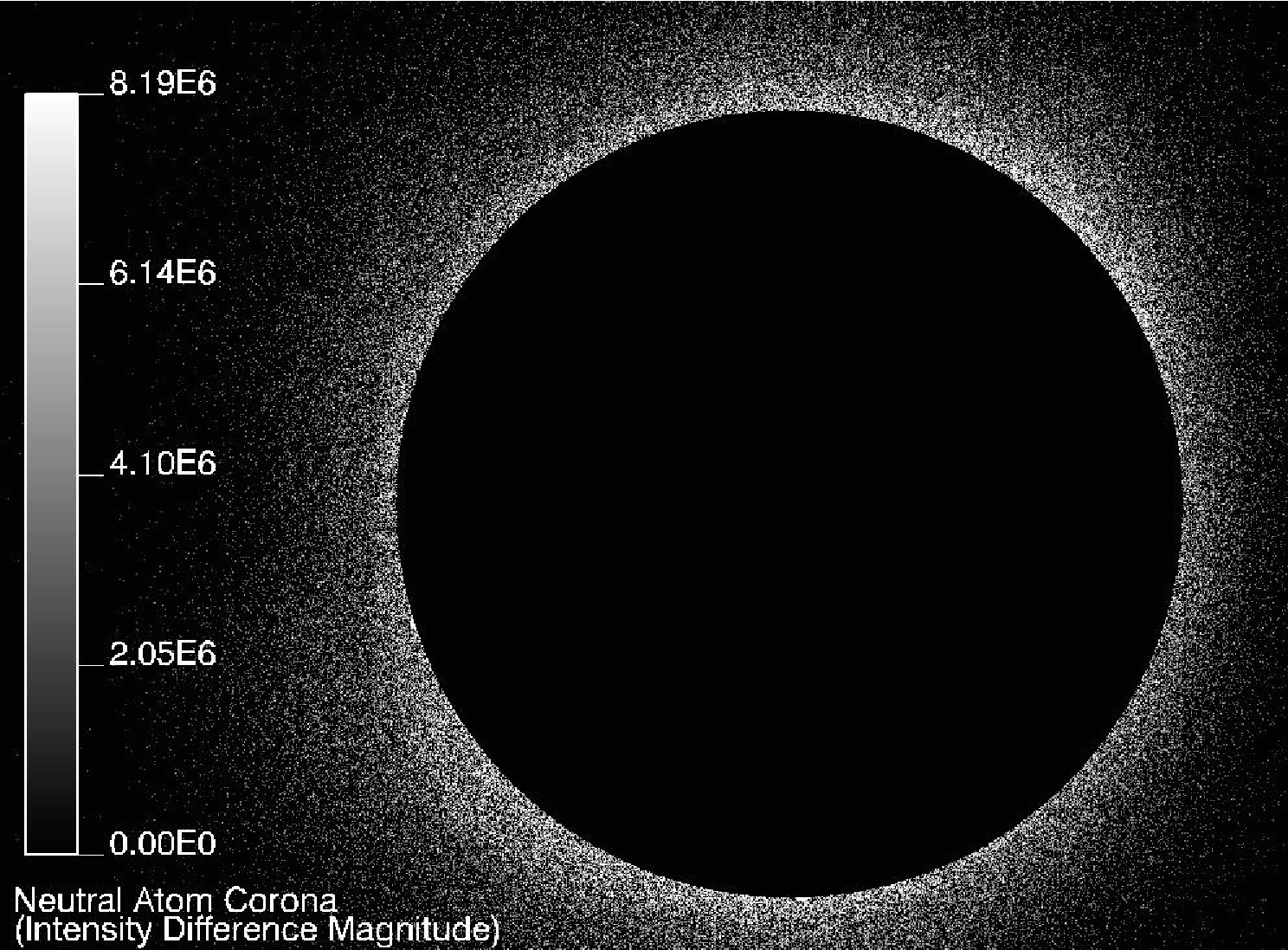}\\
\vspace{0.2cm}
\center
\includegraphics[width=7.2cm,height=5.3cm]{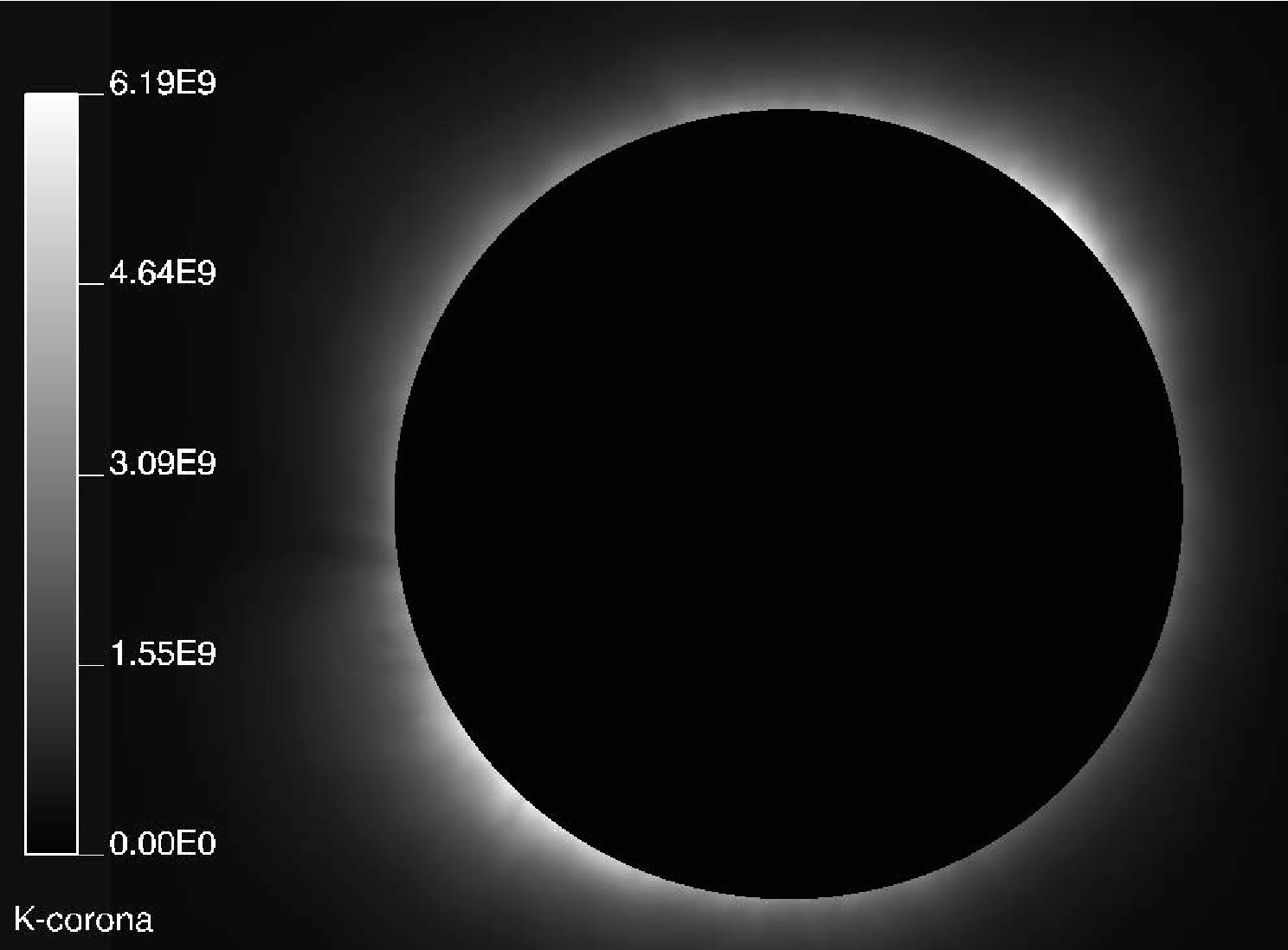}
\hspace{0.2cm}
\includegraphics[width=7.4cm,height=5.3cm]{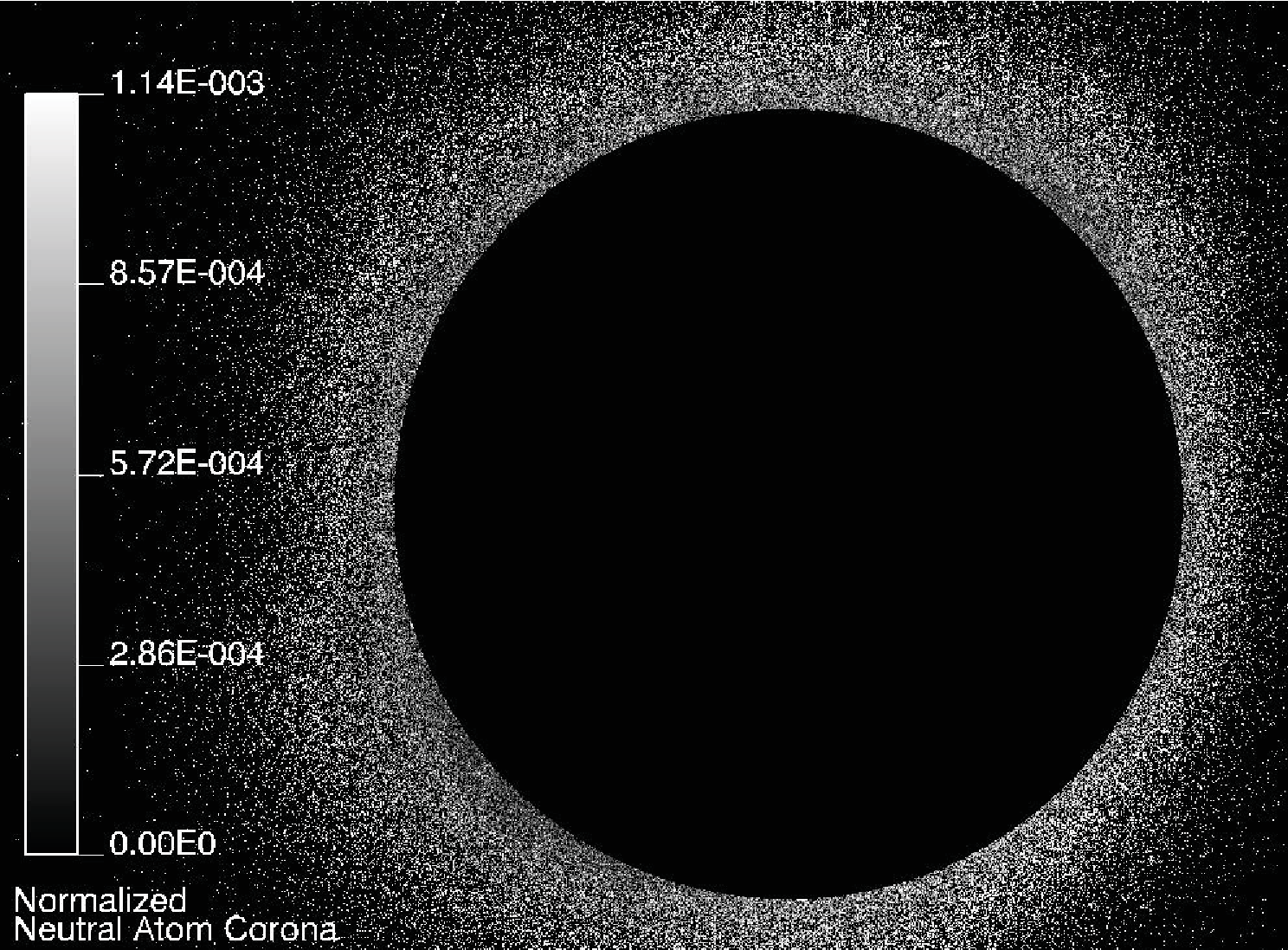}\\
\center
\vspace{0.2cm}
\includegraphics[width=7.2cm,height=5.3cm]{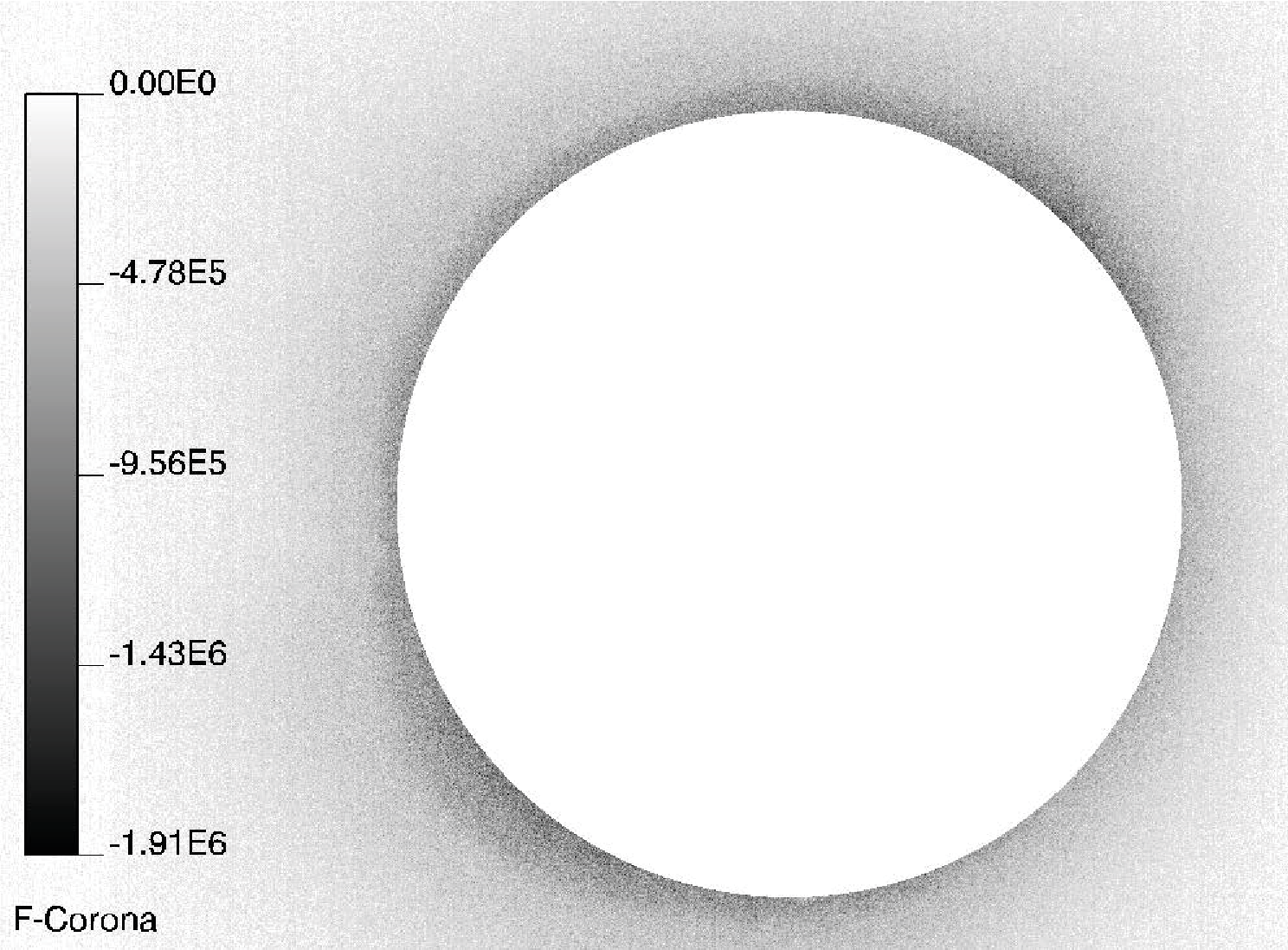}
\hspace{0.2cm}
\includegraphics[width=7.4cm,height=5.3cm]{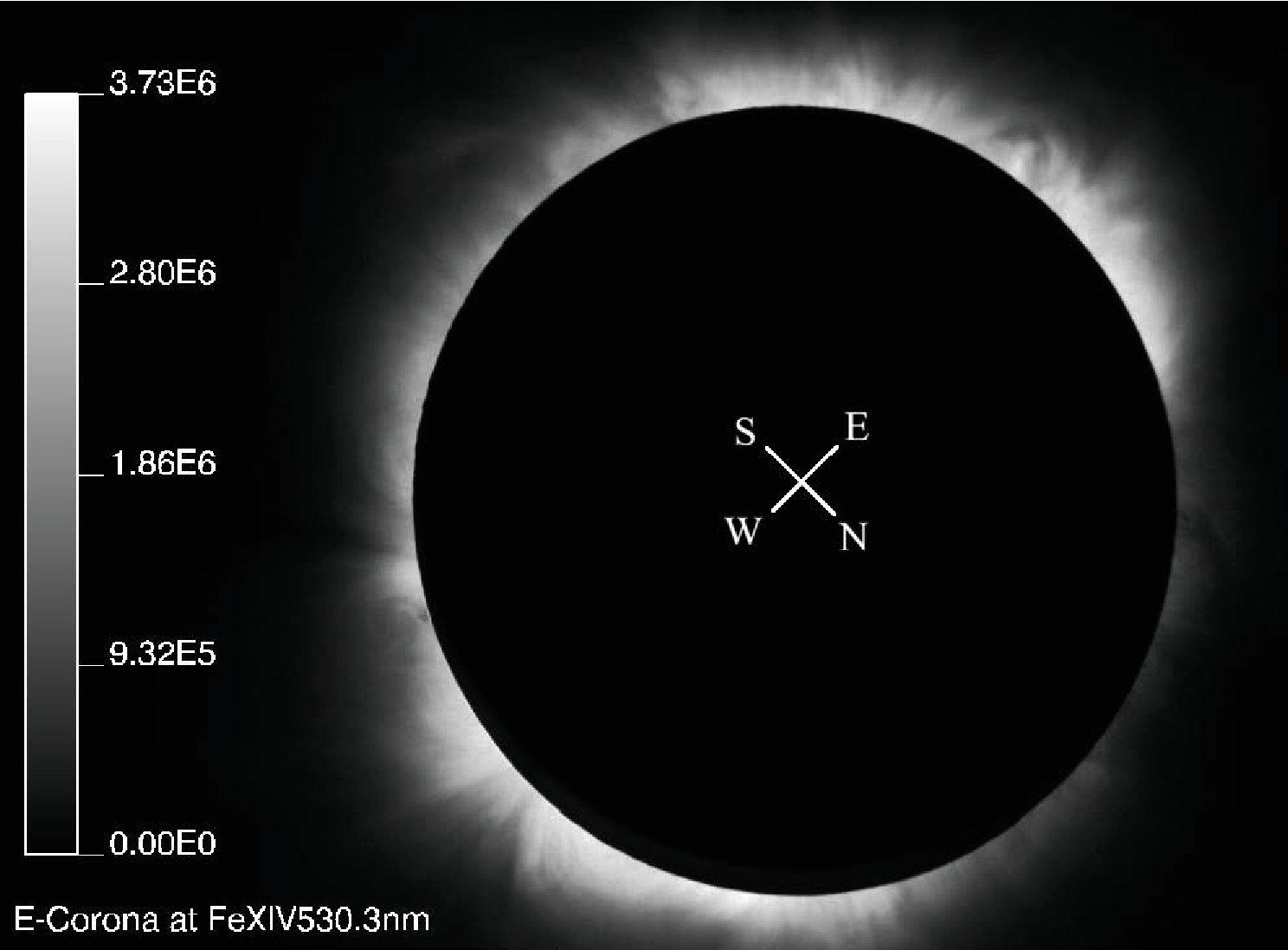}
\caption{\footnotesize Multi-faced solar corona: neutral atom
corona, inner F-corona, E-corona and K-corona. Top-left: neutral
atom corona containing both the line depression(dark) and
emission(bright), integrated over more than fifty neutral atom lines
within the observational band. Top-right: the neutral atom corona
depicted with absolute values of the intensity difference from the
adjacent continuum. Spatial points are included in these two top
maps as well as the middle right one only when their intensity
difference magnitudes are larger than three times standard
deviation(3$\sigma$) obtained in the noisy disk blocked by the moon.
Middle-left: K-corona obtained via integration over all the continua
adjacent to these neutral atom lines. Its morphology seems no change
from those `monochromatic' ones depicted in top panels of Fig.2.
Middle-right: the neutral atom corona(top-right) normalized by the
K-corona intensity. Bottom-left: inner F-corona, representing the
distribution of line depressions caused by more than fifty neutral
atom lines and thirty ion lines. Bottom-right: E-corona,
distribution of the net emission intensity of the green coronal line
above its adjacent continuum. Note that the densest distribution is
consistent in the regions of the coronal loops, especially in the
neutral atom corona plotted in the top-right panel and the F-corona.
}
\end{figure}

\section{Roles of the upward neutral atom flux in coronal heating}

The abrupt decrease of the neutral atom flux outside the coronal
loops strongly suggests neutral atom loss via ionization in the
loops. This is because that no electromagnetic forces can confine
them unless they can be heated to be ionized by collision with the
ions and then be trapped by the magnetic field via Lorentz force. In
other words, it is much easier to go outward through these coronal
loops for neutral atoms than ions. This scenario provides us a clue
of the responsibility of the outward neutral flux for the coronal
loop heating. Evidently, the above introduced asymmetries can be
reasoned to be originated from interaction of the loops with
scatterer diffusion, which can be one component of more generally
observed upward motions from solar photosphere(Tian et al., 2021).
It is noteworthy that a small fraction of neutral atoms and even
ions is witnessed to still escape from the loop-occupied layers(see
top right and bottom-right panels of Fig.3).

Actually, the above scenario can be fitted into the picture of
Cowling conductivity(Cowling,1957;  Schluter and Biermann, 1950),
resulted dominantly from collisions of neutral particles with ions
via electrodynamic coupling in presence of magnetic fields.
According to analysis by Zaitsev and Stepanov(Zaitsev and Stepanov,
2008), just a tiny concentration of the neutral atoms, e.g.,
$10^{-5}$, collided with the ions in the coronal magnetic loops, can
raise the resistance by six or more orders. Essentially, the
collisions destroy the ideal magnetohydrodynamic state into
dissipation(Klimchuk et al., 1992; Rosner et al., 1978). From the
generalized Ohm's law, it is proven that the plasma heating rate can
become more efficiently by even eight orders than that resulted from
the Spitzer resistivity in the corona(Zaitsev and Stepanov, 2008).
In fact, the same heating mechanism can be found in Tokamak device
to promote the temperature via neutral beam injection(NBI), or the
Tokamak can be regarded as the Cowling dissipation demonstration for
the coronal loop heating in laboratory on the earth(Ionson, 1984).

Now, let us give a simplified order estimation of the neutral atom
concentration according to Rybicki and Lightman(1979) after adopting
the mechanism that the free electrons with extremely high
temperature smooth out the Fraunhofer lines into the continuum as
the background radiation within the observational band. Under
assumptions that the resonant scattering is coherent and isotropic
on the scattering plane, and the coronal atmosphere is optically
thin or non-multiple scattering occurs along the line-of-sight in
the observational band, an expression is obtained of  ratio of
average density of the neutral atoms $n_{a}$ to average density of
the free electron $n_{e}$:
 \begin{equation}
 \frac{n_{a}}{n_{e}} = D_{line}\frac{\sigma_{T}}{\sigma_{a}}\frac{J_{photosphere}}{I_{c,corona}}
\end{equation}
where in $cgs$ unit system
 \begin{equation}
           \sigma_{a}=\frac{\pi e^{2}}{m_{e}c}f_{lu}\phi(\nu_{0})
 =2.654\times 10^{-2}f_{lu}\phi(\nu_{0})\hspace{0.2cm} cm^{2},  \hspace{0.5cm} \sigma_{T}=6.65\times 10^{-25}cm^{2}.
\end{equation}
In the above equations, $m_{e}$ represents the electron mass, $e$
the electron charge and $c$ the light speed in vacuum. $\sigma_{a}$
and $\sigma_{T}$ stand for respectively cross-sections of resonant
scattering by neutral atoms and that of Thompson scattering by free
electrons. $D_{line}$ represents the relative line depth, i.e., the
line depth normalized by the adjacent continuum intensity
$I_{c,corona}$. $\phi(\nu_{0})=1/\sqrt{\pi\Delta\nu_{D}}$ gives
value of line profile at the line center $\nu_{0}$ in term of
Doppler width $\Delta\nu_{D}$.  $J_{photosphere}$ indicates the
photospheric background mean radiative intensity.  The line
oscillator strength $f_{lu}$ can be found in NIST. It becomes
evident now that the distributions of the relative line depth
depicted in the middle right map of Fig.3 reflects neutral atom
concentration distributions. The smaller concentration within the
loop regions in this map may confirm the loss of the neutral atoms.

For the following estimation,  we adopt
$J_{photosphere}/I_{c,corona}=10^{6}$ at about the coronal loop
tops. Taking the relatively strong neutral iron atom lines at
522.7nm, 526.9nm and 532.8nm for example, their $D_{line}$ can reach
4.0$\%$(referred to the two lowest profiles in Fig.1). And
temperature suitable for their existence is assumed to be 6.0$\times
10^{3}K$. Their oscillator strengths read as respectively
1.23$\times 10^{-2}$, 0.84$\times 10^{-2}$ and 0.15$\times 10^{-2}$,
derived from NIST. According to the above equations, the
concentrations of these neutral iron atoms with different excitation
potentials simultaneously yielding the three Fraunhofer lines are
estimated to be respectively 2.52$\times 10^{-6}$, 1.13$\times
10^{-6}$ and 6.68$\times 10^{-6}$. Therefore, the concentration of
the neutral iron atoms detected in this observational band can
surpass 1.00$\times 10^{-5}$. Hence, it is reasonable to deduce that
the neutral atom concentration can make the Cowling dissipation
effective according to the calculations by Zaitsev and
Stepanov(2008).

Now, let us go further. As mentioned above, a tiny fraction of the
neutral atoms and once ionized ions escape still from heating in the
coronal loops as the thinnest line depression part above the
thickest. This scenario reminds us that similar physical processes
may also happen in the chromosphere and transition zone, though the
dissipation can be more easily traced in the corona due to the
'frozen effect' ascribed to much smaller ratio $\beta$  of the
plasma pressure to the magnetic one, i.e., $\beta\ll 1$. It could
also come into play in these layers below the corona with denser
neutral atoms and magnetic loops with stronger fields resulting in
$\beta\le 1$(Wang, 1993; Yalim et al., 2020). In these atmospheric
layers, smaller magnetic loops can be detected(Leake and Arber,
2006; Centeno et al., 2007; Madjarska et al., 2023; Nobrega-Siverio
et al., 2023; Judge et al., 2024a). As pointed out by
Aschwanden(2015), the heating rate is much more dependent on the
particle density than the loop length. The heating efficiencies are
also based on observational fact that emergences of 95$\%$
photospheric magnetic fluxes form the canopy tops or 'magnetic
carpet' below the corona, and only about 5$\%$ fluxes give births to
large-scale coronal loops(Priest, 2002; Rempel et al., 2014; Lites
et al., 2017). This makes the Cowling dissipation more effective in
the chromosphere(Wang, 1993; Yalim et al., 2020) and even most
efficient in the transition zone spatially related to the canopy
tops probably, thus leads to the sudden rise of temperature.

\section{Conclusions and Discussions }

We have confirmed together the presence of the neutral atom corona
and the inner F-corona primarily according to the features of the
Fraunhofer line scattering, which cannot be owned by the 'outer'
dust F-corona formed via dust scattering, let alone the telluric
atmosphere via Rayleigh scattering. Their spatial extension range is
detected within heights below half a solar radius where the dust
will be sublimated. The corresponding line emissions with more
disperse distribution are detected simultaneously and thus
strengthen the presence of the neutral atom corona. The line
depression distribution from the neutral atoms forms the primary
ingredient of the inner F-corona. According to a rough estimation,
the concentration of the neutral atoms in the inner corona can
surpass $10^{-5}$, a critical value for the Cowling dissipation
taking effect in the corona. It is noteworthy that the outer
F-corona with a  configuration of ellipsoid with a central hollow is
also superposed on the inner one discovered here due to the
projection along the line-of-sight, but their contribution is much
less due to the hollow. On the other hand, the outer F-corona
detections with higher photon collection efficiency and longer
exposure times should contain the inner F-corona as its dominant
part just above the limb due to the overlap along the line-of-sight.
This scenario can be responsible for the extra-bright intensity of
the F-corona just above the limb(say, Boe et al., 2021). This is
because that, by intuitive comparison with the full ellipsoid
configuration, the existence of the hollow with a diameter exceeding
about 2.3 solar radii in the ellipsoid configuration of the dust
F-corona cannot cause the approximately exponential increase in
intensity via scattering along the line-of-sight as approaching the
limb.

The observed hinderance of the neutral atom diffusions in the
coronal loops signifies that there should be some interactions
between the emergent neutral atoms with the ions within the coronal
loops with electric current, and the loop heating is just resulted
in via Cowling dissipation. The neutral atoms themselves will be
heated and ionized via the collisions. The origin of these neutral
atoms cannot be interpreted by Coulomb recombination of free
electrons with ions as the primary mechanism of the neutral atom
formation but the diffusion, as pointed out by Judge and
Pietarila(2004) and witnessed from those neutral atom corona maps.
This is because that the two kinds of charged particles spiral in
the loop magnetic fields with basically different radii thus
generally non-overlapped orbits, therefore probabilities of the
recombination should be much smaller than outside. In fact, the same
scenario can be responsible for the extremely low Spitzer
resistivity.

On the other hand, a tiny fraction of the neutral atoms and once
ionized ions is also observed still escaping from the coronal loops.
The leakage may be ascribed to the gaps between the magnetic loops
where no significant collisions with ions take place to hinder the
diffusion, or even a small fraction of the neutral atoms escape from
the interactions in the loops. This scenario reminds us that the
same physical process can also happen underneath the corona, so that
we can observe them in the corona via the diffusion(Judge and
Pietarila, 2004). In order to stop the escape of neutral atoms in
the underneath atmospheric layers into the corona according to the
presently proposed mechanism, the neutral atoms should completely
ionized in the transition zone due to no gaps between the magnetic
canopies and heating efficiency should reach 100$\%$ within the
loops, or other mechanisms prevent the escape. However, no
observations have been found to support the view or none of theory
supports it to our knowledge, probably because electromagnetic
forces provide more confinement on diffusions of the free electrons
and ions from the underneath layers into the corona. In fact, these
neutral atoms with much low temperatures can be accounted for as one
part in the low energy wing of particle distribution curve of much
higher temperatures, though the distribution could be departed
greatly from the Maxwellian one.

Efficiency of the Cowling dissipation caused by the outward neutral
atom fluxes depends also on global abundance of the coronal loops,
which have a configuration simpler than that needed for magnetic
reconnections, as witnessed in the E-corona map. They may indicate
different heating phases. Therefore, globality of the Cowling
dissipation can be the most efficient mechanism for global coronal
heating. On the other hand, intermittent emergences of magnetic loop
into the corona generated via dynamo mechanism(Mok et al., 2008;
Martinez et al., 2009; Judge and Kuin, 2024b) guarantee the
compensation of the coronal loops dissipated via radiation and
transverse conduction to heat the surroundings(Litwin and Rosner,
1993). That the ubiquitous convection driving the loop feet in the
convection zone induce electric currents within the loops by
Faraday's law of electromagnetic induction provides the energy
sources for the dissipation. such a scenario reflects the
self-regulation of coronal heating similar to the small-scale
reconnections(Uzdensky, 2007). This shows a wonder in the Sun and
supplies us insight into the solar inorganic life.

As pointed out previously, the Cowling dissipation plays a role as a
trunk in a unified scenario about the primary heating cascades along
atmospheric height from the chromosphere through the transition zone
to the corona as $\beta$ consistently decreases. This will present
one example of coupling among all these atmospheric layers(Ryutova
and Shine, 2006). This view is based on not only origin of the
neutral outward fluxes deduced here but also the solid observations
that emergences of dominant photospheric magnetic fluxes braid the
canopies or 'magnetic carpet' just below the corona, only remnant
fluxes survive outwards as large-scale coronal loops(Priest et al.,
2002; Magara, 2012; Rempel, 2014; Lites et al., 2017) via diffusion.
These observational facts can make the Cowling dissipation more
effective for the heating in the atmospheric layers just beneath the
corona with denser particles and magnetic loops and cause the rapid
rise of the temperature in the transition zone.

Besides the coronal loops, current sheets in separatrices(e.g.,
Priest et al., 2002) provide another platform for the Cowling
dissipation. The neutral fluxes can permeate into the current sheets
to promote crucially the Joule dissipation at the beginning of
magnetic reconnections when they are not ionized, thus enhance
considerably the heating efficiency(Uzdensky, 2007; Ni et al., 2012;
Yalim et al., 2020; Wargnier et al., 2023). Microflares and
nanoflares with energy release in the current sheets were proposed
by Parker(1972; 1988) to be a main mechanism for the coronal
heating. Since then, it was developed by Priest and his
cooperators(say, Priest et al., 2002; Priest, 2011) via a coronal
tectonics model. Otherwise, the dissipation can also take effect in
other small scale magnetic reconnection regions like
spicules(Samanta et al., 2019) and Ellerman bombs(Fang et al.,
2006). Finally, the heating efficiency may be greatly improved by
energy dissipation of magnetohydrodynamic(MHD) waves, as a main
mechanism proposed by many researchers, e.g., van Ballegooijen et
al.(2017). Some of these waves can be generated by impact of the
neutral atom upflows with the coronal loops that are sensitive wave
resonators(Zaitsev and Stepanov, 2008; Priest, 2011). The dark
curved front, seen in the neutral atom corona in the middle right
panel of Fig.3, may provide us a trace of their existence.
Therefore, the neutral flux forming the neutral atom corona becomes
a missed but crucial ingradient in both the resistive and reactive
ways of the coronal heating shown in Fig.2.4 of Judge and
Ionson(2024).

\acknowledgments{{\bf Acknowledgements} This work is sponsored by
the National Key R$\&$D Program of China under the grant number
2021YFA1600502(2021YFA1600500), National Science Foundation of China
(NSFC) under the grant numbers 11527804. }

\centering{\bf References}

\raggedright{

Allen, C.W., The spectrum of the corona at the eclipse of 1940
October I, {\it Monthly Notes of Royal Society}, {\bf 106},
137-150(1946)

Aschwanden, M.J., Physics of the solar corona, {\it Springer,
Published in association with Praxis Publishing Chichester, UK},
{ISBN: 3-540-22321-5}, 2015

van Ballegooijen, A.A.,  Asgani-Targhi, M. and Voss, A., The heating
of solar coronal loops by Alfv$\acute{e}$n wave turbulence. {\it the
Astrophysical Journal}, {\bf 849}:46(23pp)(2017)

Bazin, C.  and Koutchmy, S., Helium shells and faint emission lines
from slitless flash spectra. {\it Journal of Advanced Research},
{\bf 4}, 307-313(2013)

Blackwell, D.E., and Petford, A.D., Observations of the 1963 July 20
solar eclipse I. Spectroscopic separation of the F and K components
of the solar corona at large distances from the sun, {\it Monthly
Notes of Royal Astronomical Society}, {\bf 131}, 383-398(1966a)

Blackwell, D.E., and Petford, A.D., Observations of the 1963 July 20
solar eclipse II. The electron density in the solar corona in the
region 5$<R/R_{0}<16$ obtained from measurements of Fraunhofer line
depth, and the polarization of the F corona, {\it Monthly Notes of
Royal Astronomical Society}, {\bf 131}, 383-398(1966b)

Boe, B., Habbal, S.,Downs,  C. and Druckmueller, M., The color and
brightness of the F-corona inferred from the 2019 July 2 total solar
eclipse. {\it the Astrophysical Journal}, {\bf 912}:44(15pp)(2021)

Burtovoi, A. , Naletto, G.,  Dolei, S., Spadaro, D., Romoli, M.,
Landini,  F. and De Leo,  Y.,  Measuring the F-corona intensity
through time correlation of total and polarized visible light
images. {\it Astronomy $\&$ Astrophysics}, {\bf 659}, A50(16
pages)(2022).

Centeno, R.,  Socas-Navarro, H.,  Lites, B., et al., 2007, Emergence
of Small-Scale Magnetic Loops in the Quiet-Sun Internetwork, {\it
the Astrophysical Journal}, {\bf 666}, L137-L140

Cowling, T.G., {\it Magnetohydrodynamics}, Interscience Publishers,
New York, 1957

Deutsch, A.J., and Righini, G., An Airborne Observation of the
Coronal Spectrum at the Eclipse of July 20, 1963, {\it the
Astrophysical Journal}, {\bf 140}, 313-318(1964)

Edl$\acute{e}$n, B., Die Deutung der Emissionslinien im Spektrum der
Sonnenkorona. Mit 6 Abbildungen. {\it Zeitschrift fur Astrophysik},
{\bf 22}, 30-64(1943).

Fang, C. , Tang, Y. H.,  Xu, Z.,  et al., Spectral Analysis of
Ellerman Bombs. {\it the Astrophysical Journal}, {\bf 643},
1325-1336(2006).

Grotrian, Von W., $\ddot{U}$ber das Fraunhofersche Spektrum der
Sonnenkorona. {\it Zeitschrift f$\ddot{u}$r Astrophysik}, {\bf 8},
124-137(1934).

Ionson, J.A.,  A unified theory of electrodynamic coupling in
coronal magnetic loops: the coronal heating problem. {\it
Astrophysical Journal}, {\bf 276}, 357-368(1984).

Judge,  P.J. and Ionson, J.A. ,  The problem of coronal heating,
2024,  {\it Astrophysics and Space Science Library}, Volume 470,
Springer, ISBN 978-3-031-46272-6

Judge, P.G. and Kuin, N.P.M., On the intermittency of hot plasma
loops in the solar corona. {\it the Astrophysical Journal}, {\bf
970}:130(9pp)(2024).

Judge, P.G. and Pietarila, A., On the formation of
extreme-ultraviolet helium lines in the sun: analysis of SOHO data.
{\it the Astrophysical Journal}, {\bf 606}, 1258-1275(2004).

Klimchuk, J.A., Lemen, J.R., Feldman, U.,  et al., Thickness
Variations along Coronal Loops Observed by the Soft X-Ray Telescope
on YOHKOH. {\it Publications of the Astronomical Society of Japan},
{\bf 44}, L181-L185(1992).

Koutchmy, S. and Magnant, F., On the observation of the F-corona in
the vicinity of the solar limb, {\it Astronomy $\&$ Astrophysics},
{\bf 632}, A86(11-14)(2019)

Koutchmy, S., Baudin, F., Abdi, Sh.  et al., New deep coronal
spectra from the 2017 total solar eclipse, {\it the Astrophysical
Journal}, {\bf 186}, 671-677(1973)

Kraus, I., Bourdin, Ph.-A., Zender, J., et al., 'Coronal bright
point statistics: II. Magnetic polarities and mini loops', {\it
Astronomy $\&$ Astrophysics}, {\bf 694}, A240(7pp)

Kuhn,  J.R., Arnaud, J., Jaeggli, S.,  Detection of an Extended
Near-Sun Neutral Helium Cloud from Ground-based Infrared Coronagraph
Spectropolarimetry. {\it the Astrophysical Journal}, {\bf 667},
L203-L205(2007).

Lamy, P.L.,  Gilardy, H. and Liebaria, A., Observations of the solar
F-corona from space. {\it Space Sci. Rev.}, {\bf 218}, 53(72
pages)(2022).

Leake J.E. and Arber, T.D., The emergence of magnetic flux through a
partially ionised solar atmosphere. {\it Astronomy $\&$
Astrophysics}, {\bf 450}, 805-818(2006)

Lites, B.W., Rempel, M., Borrero, J.M., et al., Are internetwork
magnetic fields in the solar photosphere horizontal or vertical?.
{\it the Astrophysical Journal}, {\bf 412}, 14-20(2017).

Litwin, C. and Rosner, R., On the structure of solar and stellar
coronae: loops and loop heat transport. {\it the Astrophysical
Journal}, {\bf 412}, 375-385(1993).

Madjarska, M.,Galsgaard, K.  and Wiegelmann, T., Photospheric
magnetic flux and coronal emission properties of small-scale bright
and faint loops in the quiet sun. {\it Astronomy $\&$ Astrophysics},
{\bf 678}, A32-A38(2023).

Magara, T., How Much does a magnetic flux tube emerge into the Solar
atmosphere?. {\it the Astrophysical Journal}, {\bf 748},
53(7pp)(2012).

Martinez Gonzalez, M. J. $\&$ Bellot Rubio, L. R., 2009, Emergence
of Small-scale Magnetic Loops Through the Quiet Solar Atmosphere,
{\it the Astrophysical Journal}, {\bf 700}, 1391-1403

Menzel, D.H. and Pasachoff, J.M., On the obliteration of strong
Fraunhofer lines by electron scattering in the solar corona, {\it
Publications of Astronomical Society of the Pacific}, {\bf 80},
458-461(1968)

Moise, E., Raymond, J.  and Kuhn, J. R., Properties of the diffuse
neutral helium in the inner heliosphere. {\it the Astrophysical
Journal}, {\bf 722}:1411-1415(2010).

Mok, Y.; Mikic, Z.; Lionello, R., et al., 2008, The Formation of
Coronal Loops by Thermal Instability in Three Dimensions, {\it the
Astrophysical Journal}, {\bf 679}, L161-L165

Morgan, H. and Habbal, S.R., The long-term stability of the visible
F corona at heights of 3-6 R$_{\bigodot}$. {\it Astronomy $\&$
Astrophysics},  {\bf 471}, L47-50(2007).

Ni Lei, Roussev, I.I., Lin, Jun, Ziegler, U., Impact of
temperature-dependent resistivity and thermal conduction on plasmoid
instabilities in current sheets in the solar corona. {\it the
Astrophysical Journal}, {\bf 758}, 20(11pp)(2015).

Nobrega-Siverio, D.,Moreno-Insertis,  F., Galsgaard, K. ,Krikova, K.
and  van der Voort, L.P., Deciphering solar coronal heating:
Energizing small-scale loops through surface convection. {\it the
Astrophysical Journal}, {\bf 958}, L38(8pp)(2023).

Parker, E.N., Topological dissipation and the small-scale fields in
turbulent gases. {\it the Astrophysical Journal}, {\bf 174},
499-510(1972).

Parker, E.N., Nanoflares and the solar X-ray corona. {\it the
Astrophysical Journal}, {\bf 330}, 474-479(1988).

Priest, E.R., Heyvaerts, J.F.,  Title, A.M., A flux-tube tectonics
model for solar coronal heating driven by the magnetic carpet. {\it
the Astrophysical Journal}, {\bf 576}, 533-551(2002).

Priest, E.R., The flux tube tectonics model for coronal heating,
{\it Journal of atmospheric and solar-terrestial physics}, {\bf 73},
271-276(2011).

Qu, Z.Q.,  A Fiber Arrayed Solar Optical Telescope. {\it Solar
Polarization 6, ASP Conference Series}, Vol.437,423-431(2011), Kuhn,
Berdyugina, Harrington, Keil, Lin, Rimmele, and Trujillo-Bueno, eds.

Qu, Z.Q.,  Chang, L.,  Cheng, X. M.,  et al., Prototype FASOT. {\it
Solar Polarization 7}, ASP Conference Series, Vol.489,
263-270(2014), K. N. Nagendra, J. O. Stenflo, Zhongquan Qu, and M.
Sampoorna, eds.

Qu, Z.Q.,  Chang, L., Dun,  G.T.,  Cheng, X.M.,  et al.,
Spectropolarimetry of Fraunhofer lines in local upper solar
atmosphere. {\it the Astrophysical Journal}, {\bf 974},
63(13pp)(2024).

Rempel, M., Nonlinear simulations of quiet sun magnetism: on the
contribution from a small-scale dynamo. {\it the Astrophysical
Journal}, {\bf 789}, 132-139(2014).

Rosner, R., Golub, L., Coppi, B. and Vaiana, G. S., Heating of
coronal plasma by anomalous current dissipation. {\it the
Astrophysical Journal}, {\bf 222}, 317-332(1978).

Russell, H.N.,  On meteoric matter near the stars. {\it the
Astrophysical Journal}, {\bf 69},49-71(1929).

Rybicki, G.B., $\&$ Lightman, A.P., 1991, {\it Radiative Processes
in Astrophysics}, John Wiley $\&$ Sons

Ryutova, M.  and Shine, R., Coupling effects throughout the solar
atmosphere: Emerging magnetic flux and structure formation. {Journal
of Geophysical Research}, {\bf 111}, A03101(10pp)(2006).

Samanta, T., Tian, Hui, Yurchyshyn, V., et al., Generation of solar
spicules and subsequent atmospheric heating. {\it Science}, {\bf
366}, 890-894(2019).

Schluter, A.  and Biermann, L.Z., Interstellare magnetfelder. {\it
Naturforsch}, {\bf A5}, 237(1950).

Stellmacher, G.  and Koutchmy, S.,  Study of low dispersion eclipse
spectra: Observation of weak low extinction emission lines in the
corona. {\it Astronomy $\&$ Astrophysics}, {\bf 35}, 43-48(1974).

Tian, H., Hara,  L., Baker, D.,  et al., Upflows in the upper solar
atmosphere. {\it Solar Physics}, {\bf 296}, 47(2021).

Title, A.M., AIA team, The Atmospheric Imaging Assembly on the Solar
Dynamics Observatory, American Astronomical Society, {\it SPD
meeting} No.37, id.36.05; Bulletin of the American Astronomical
Society, Vol. 38, pp.261-274(2006)

Uzdensky, D.A., Self-regulation of solar coronal heating process via
the collisionless reconnection condition. {\it Physical Review
Letters}, {\bf 99}, 261101(2007).

Wang Jingxiu, Electric conductivity of lower solar atmosphere. {\it
ASP Conference Series}, {\bf 46}, 465-468(1993).

Wargnier, Q.M., Martinez-Sykora, J., Hansteen, V.H.  andDe Pontieu,
B.,  Multifluid simulations of upper-chromospheric magnetic
reconnection with helium-hydrogen mixture. {\it the Astrophysical
Journal}, {\bf 946}:115(24pp)(2023).

Yalim,  M.S.,  Prasad, A., Pogorelov, N.V., et al., Effects of
Cowling resistivity in the weakly ionized chromosphere. {\it the
Astrophysical Journal}, {\bf 899}: L4(7pp)(2020).

Zaitsev, V.  and Stepanov, A.V.,  Coronal magnetic loops. {\it
Physics-Uspekhi}, {\bf 51}, 1123-1160(2008).

}
\end{document}